\documentclass[fleqn,usenatbib]{mnras}

\usepackage{longtable}
\usepackage{booktabs}
\pdfoutput=1

\usepackage[T1]{fontenc}

\usepackage{setspace}

\usepackage{graphicx}	
\usepackage{amsmath}	
\usepackage{amssymb}	
\usepackage{newtxtext,newtxmath}

\usepackage{xcolor}
\definecolor{pink}{rgb}{0.98, 0.38, 0.5}

\newcommand{\Vr}{$V_r$}
\newcommand{\Vphi}{$V_{\phi}$}
\newcommand{\Vtheta}{$V_{\theta}$}
\newcommand{\mstar}{$\rm M_{\star}$}
\newcommand{\mstartrue}{$\rm M_{\star}^{true}$}
\newcommand{\mstarest}{$\rm M_{\star}^{est}$}
\newcommand{\feh}{$\langle\rm [Fe/H]\rangle$}
\newcommand{\fehsimall}{$\rm [Fe/H]_{sim}$}
\newcommand{\fehsimwindow}{$\rm [Fe/H]_{sim}^\ast$}
\newcommand{\fehsimshifted}{$\rm [Fe/H]_{sim}^{\ast\downarrow}$}

\newcommand{\fehgrad}{$\rm \nabla [Fe/H]_{GES}^\ast$}
\newcommand{\logmassratio}{$\rm log_{10}(M_{\star}^{est}/M_{\star}^{true})$}

\title[Different selections of GES stars]{Can we really pick and choose? Benchmarking various selections of Gaia Enceladus/Sausage stars in observations with simulations}

\author[A. Carrillo et al.]{
Andreia Carrillo$^{1,2}$\thanks{E-mail: andreia.carrillo@durham.ac.uk},
Alis J. Deason$^{1,2}$, Azadeh Fattahi$^{1}$, Thomas M. Callingham$^{3}$, Robert J. J. Grand$^{4,5,6}$
\\
$^{1}$Institute for Computational Cosmology, Department of Physics, Durham University, Durham DH1 3LE, U.K \\
$^{2}$Centre for Extragalactic Astronomy, Department of Physics, University of Durham, South Road, Durham DH1 3LE, UK
\\
$^{3}$Kapteyn Astronomical Institute, University of Groningen, Landleven 12, 9747 AD Groningen, The Netherlands\\
$^4$Astrophysics Research Institute, Liverpool John Moores University, 146 Brownlow Hill, Liverpool, L3 5RF, UK\\
$^{5}$
Instituto de Astrofisica de Canarias, Calle Via Lactea s/n, E-38205 La Laguna, Tenerife, Spain\\
$^{6}$Departamento de Astrofisica, Universidad de La Laguna, Av. del Astrofisico Francisco Sanchez s/n, E-38206, La Laguna, Tenerife, Spain
}

\date{Accepted XXX. Received YYY; in original form ZZZ}

\pubyear{2015}

\begin{document}
\label{firstpage}
\pagerange{\pageref{firstpage}--\pageref{lastpage}}
\maketitle

\begin{abstract}

Large spectroscopic surveys plus Gaia astrometry have shown us that the inner stellar halo of the Galaxy is dominated by the debris of Gaia Enceladus/Sausage (GES). With the richness of data at hand, there are a myriad of ways these accreted stars have been selected. We investigate these GES selections and their effects on the inferred progenitor properties using data constructed from APOGEE and Gaia. We explore selections made in eccentricity, energy-angular momentum (E-Lz), radial action-angular momentum (Jr-Lz), action diamond, and [Mg/Mn]-[Al/Fe] in the observations, selecting between 144 and 1,279 GES stars with varying contamination from in-situ and other accreted stars. We also use the Auriga cosmological hydrodynamic simulations to benchmark the different GES dynamical selections. Applying the same observational GES cuts to nine Auriga galaxies with a GES, we find that the Jr-Lz method is best for sample purity and the eccentricity method for completeness. Given the average metallicity of GES (-1.28 < [Fe/H] < -1.18), we use the $z=0$ mass-metallicity relationship to find an average \mstar~of $\sim 4 \times 10^{8}$ $\rm M_{\odot}$.  We adopt a similar procedure and derive \mstar~for the GES-like systems in Auriga and find that the eccentricity method overestimates the true \mstar~by $\sim2.6\times$ while E-Lz underestimates by $\sim0.7\times$. Lastly, we estimate the total mass of GES to be $\rm 10^{10.5 - 11.1}~M_{\odot}$ using the relationship between the metallicity gradient and the GES-to-in-situ energy ratio. In the end, we cannot just `pick and choose' how we select GES stars, and instead should be motivated by the science question.

\end{abstract}




\section{Introduction}


Galaxies grow from the accretion of other smaller galaxies and the Milky Way is no stranger to this process. Particularly, our Galaxy has had a relatively quiet recent merger history \citep{gilmore02}, save for the current interactions with Sagittarius \citep{ibata94,majewski17} and the Magellanic Clouds \citep{gomez15,laporte18}, in addition to dwarf galaxies and globular clusters that are tidally disrupted into streams in the stellar halo (e.g., \citealt{belokurov06, shipp18}). However, a seemingly significant merger in hiding has been hinted at in the past \citep{nissen97,chiba00,nissen10} because of the distinct kinematics and chemistry imprinted in some halo stars in the local Solar neighborhood. With data from the Gaia mission providing astrometry for $> 10^9$ stars, in tandem with large spectroscopic surveys providing chemistry for $> 10^6$ stars, we have indeed confirmed the existence of this merger we now call Gaia-Enceladus/Sausage (GES; \citealt{belokurov18}; \citealt{helmi18}). With virial mass, $\rm M_{vir} > 10^{10} M_{\odot}$ \citep{belokurov18}, GES has undoubtedly changed the course of the evolution and the resulting picture of the Milky Way---from dominating the inner stellar halo and causing a break in the stellar halo density profile \citep{carollo10,deason18,han22}, to heating up pre-existing disc stars to halo kinematics \citep{belokurov20, bonaca20}, to potentially diluting the metallicity in the interstellar medium (ISM) causing the distinct chemistry of the Milky Way thin and thick discs \citep{grand20,ciuca22}. The role of GES in understanding the formation and evolution of the Milky Way is therefore unmistakably important.  

An obvious next step is putting GES into context with other intact and disrupted satellites (e.g., \citealt{fattahi20}; \citealt{hasselquist21}; \citealt{naidu22}). Many spectroscopic surveys such as the Apache Point Observatory Galactic Evolution Experiment (APOGEE; \citealt{majewski17}) and the Hectochelle in the Halo at High
Resolution Survey (H3 Survey; \citealt{conroy19}) have been extremely powerful for these purposes. A glaring theme from these works however is that the stellar halo is comprised of substructures. The extended nature of GES in its kinematic, dynamical, and chemical properties, especially makes it more prone to overlap with accreted stars from other progenitors or even in-situ populations. On the other hand, distinct phase-space substructures in the halo such as Wukong, Arjuna, I'itoi, and Sequoia have been suggested to be actually part of GES (e.g., \citealt{naidu20}; \citealt{horta22}). It is becoming increasingly apparent and important to first understand which stars belong to GES and which do not, as this would grossly affect the properties that we infer for its progenitor.

This is further complicated by the fact that different surveys with different selection functions \textit{also} have different ways of selecting the GES population. 
This, therefore, raises the question \textit{"can we really pick and choose the way we select GES stars?''} One of the main ways GES stars have been selected is in E-Lz space as originally done in \citet{helmi18} with Gaia DR2 and APOGEE DR14. Here they find that GES forms a separate sequence in the colour-magnitude diagram and has slightly retrograde kinematics. Another widely used selection is in eccentricity space. A high eccentricity cut i.e., $e >$ 0.7, is made to select GES stars that have a large dispersion in \Vr~compared to \Vphi~(e.g., \citealt{mackereth19} with APOGEE; \citealt{naidu20} with H3), therefore falling on the \textit{"sausage''} region in the \Vphi-\Vr~diagram \citep{belokurov18}. A selection in Jr-Lz was also introduced by \citet{feuillet20} using data from the SkyMapper Southern Sky Survey \citep{casagrande19} and Gaia DR2. Specifically, they selected stars with high radial action and found that this is a cleaner selection as showcased by a normal, Gaussian metallicity distribution function (MDF) for the resulting GES sample. Multiple works have also used the \textit{action diamond} to select GES stars (e.g., \citealt{myeong19}; \citealt{lane22}) which uses the actions $J_R$, $J_z$, and $J_{\phi}$ (or $L_z$), and delineates stars that have prograde vs retrograde and polar vs radial orbits. Others have used APOGEE data and selected purely on chemistry specifically in [Mg/Mn] vs [Al/Fe], as this distinctly separates the accreted from the in-situ stars \citep{hawkins16, das20, carrillo22a}. Specifically, this selection benefits from chemical abundances being an intrinsic and inherited property of stars which could then be used to tag them to different populations \citep{freeman02}. 

The variety of selection methods for GES stars necessitates comparisons of these different selections and the resulting GES samples. \citet{buder22} used abundances from the Galactic Archaeology HERMES (GALAH) DR3 and found that the GES chemical selection made in [Mg/Mn] vs [Na/Fe] space overlaps with 29\% of their dynamical selection made in Jr-Lz.  \citet{lane22} explored six kinematic selections for GES in an idealized simulation where they found that the scaled action space is best in separating GES from the isotropic halo and achieves sample purity of 82\%. \citet{limberg22} found a similar level of sample purity for their GES sample using the Jr-Lz selection in observations, estimating the contamination from APOGEE abundances. 

Selecting GES stars is non-trivial as the performance of these selections are usually evaluated based on cross-validations between kinematics and chemistry. While this is the best that we can do in the observations, one should be cognizant that this is far from perfect. For example, the samples are overlapping but still different for the Milky Way thin and thick discs depending on how we define them---chemically, kinematically, or spatially.   


This is where simulations come in, as they can help us understand the biases and contamination from the different GES selection methods since we \textit{know} which stars truly come from the GES-like system. Indeed, Milky Way-like galaxies that have accreted GES-like systems have been found in cosmological simulations. \citet{mackereth19} used EAGLE simulations and found that the GES in the observations are well-reproduced by progenitors with $ 10^{8.5}\lesssim$~\mstar~$\rm  \lesssim 10^{9}~M_{\odot}$. \citet{fattahi19} also explored the Auriga hydrodynamical cosmological zoom-in simulations \citep{grand17} and found halos with radially anisotropic stellar halo populations at higher metallicities, similar to GES. These are produced by progenitors with $ 10^{9}\lesssim$~\mstar~$\rm  \lesssim 10^{10}~M_{\odot}$ that merged with the host galaxy $6-10$ Gyr ago. The properties of these halos have been followed up in detail by \citet{orkney23} where they found a diversity in the GES progenitors in Auriga, for example in their diskiness, or in their associated satellite population. These works show that the simulations are extremely valuable in understanding the nature of GES.  

In this paper, we aim to understand how the different GES selections compare to each other by systematically selecting from a sample constructed from APOGEE DR17 and Gaia DR3, and applying the same selections to Milky Way-GES systems in the Auriga simulations. In particular, we investigate the E-Lz, eccentricity, and Jr-Lz selections in both observations and simulations, and the [Mg/Mn] vs [Al/Fe] selection in the observations. We specifically aim to answer the following questions: (1) What is the overlap between each selection and what contamination do we have from other accreted and in-situ components? (2) What progenitor properties do we infer from these different selections? (3) What selection is the purest and what selection is the most complete? In trying to address these questions, we want to impart more careful deliberation on how we select GES stars, as this would affect the sample data and ultimately the nature and picture of the GES progenitor that we paint from this.    

This work is organised as follows: In Section \ref{sec:observation_data} we describe our observational data and the selections made within. In Section \ref{sec:chemical_vs_kinematic_obs} we compare the different GES samples and their inferred progenitor properties. Section \ref{sec:simulation_data} focuses on the Milky Way-GES systems in the Auriga simulations, applying the selections we use in the observations, and investigating the performance of these selections. In Section \ref{sec:discussion} we discuss and explore other avenues of validating the GES selections, and lastly, we summarize the results from this work in Section \ref{sec:conclusions}. 

\section{Accreted stars from an observational perspective}
\label{sec:observation_data}



In this section, we construct a chemodynamical data set from the Gaia Data DR3 \citep{gaiaedr3} and APOGEE DR17 \citep{apogeedr17}. We take the positions, radial velocity, and proper motions from Gaia DR3 as well as the Gaia Early DR3-derived distances from \citet{bailerjones21} for calculating dynamical properties of the sample. 

The APOGEE data, taken in the \textit{H}-band (1.5-1.7 $\mu$m) with R $\sim$ 22,500, contains individual element abundances for up to 20 species for 733,901 stars. Of the elements available, we specifically use the Mg, Mn, and Al which are useful for selecting accreted stars purely from a chemical standpoint (e.g., \citealt{hawkins15}, \citealt{das20}, \citealt{carrillo22a}). 

To ensure the quality of our sample, we make the following cuts in this crossmatched data set: 
\begin{itemize}
    \item{parallax error / parallax $<$ 0.20}
    \item{radial velocity error $<$ 2.0 km $\rm s^{-1}$}
    \item{STARFLAG = 0 \footnote{for good spectra} and ASPCAPFLAG = 0 \footnote{for converged stellar parameters}}
    \item{[Mg/Fe], [Mn/Fe], [Al/Fe], and [Fe/H] \texttt{are not null}}
\end{itemize}

To reduce the contamination from in-situ material, we also avoid the disc and apply a cut of |z|$>$1 kpc. With these cuts, we have a resulting sample of 41,534 stars. We use the \citet{cautun20} potential within the \texttt{galpy} package \citep{bovy15} to derive the orbital properties of this Gaia-APOGEE crossmatched sample. We discuss later in detail in Section \ref{sec:kinematic_obs} the effects of observational uncertainities on the derived properties. From the parent sample of 41,534 stars, we select GES through various ways, as will be discussed next.

\subsection{Chemical selection of accreted stars in observations}
\label{sec:chemical_obs}


\begin{figure}
 \center
\includegraphics[width=0.9\columnwidth]{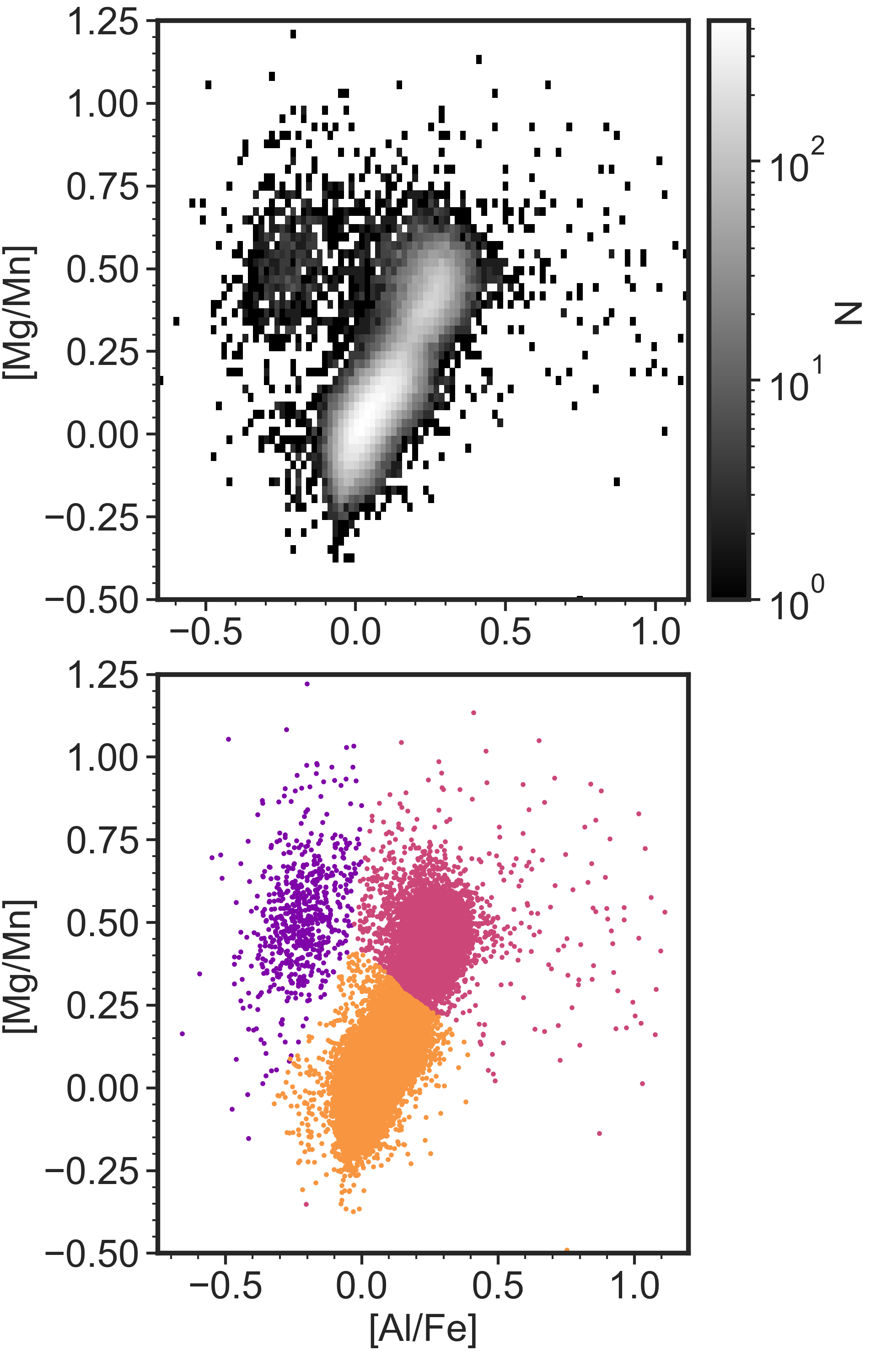}
 \caption{[Mg/Mn] vs [Al/Fe] from APOGEE DR17 used for the chemical selection of accreted stars. \textit{Top panel}: 2D histogram showing the distinct group of stars at [Al/Fe] <0 and and [Mg/Mn] > 0.25 that corresponds to accreted stars. \textit{Bottom panel}: Three separate regions in the diagram that roughly correspond to accreted stars (Region 1, purple), thick disc (Region 2, pink), and thin disc (Region 3, orange).}
 \label{fig:mgmnal}
\end{figure}

\begin{figure}
 \includegraphics[width=\columnwidth]{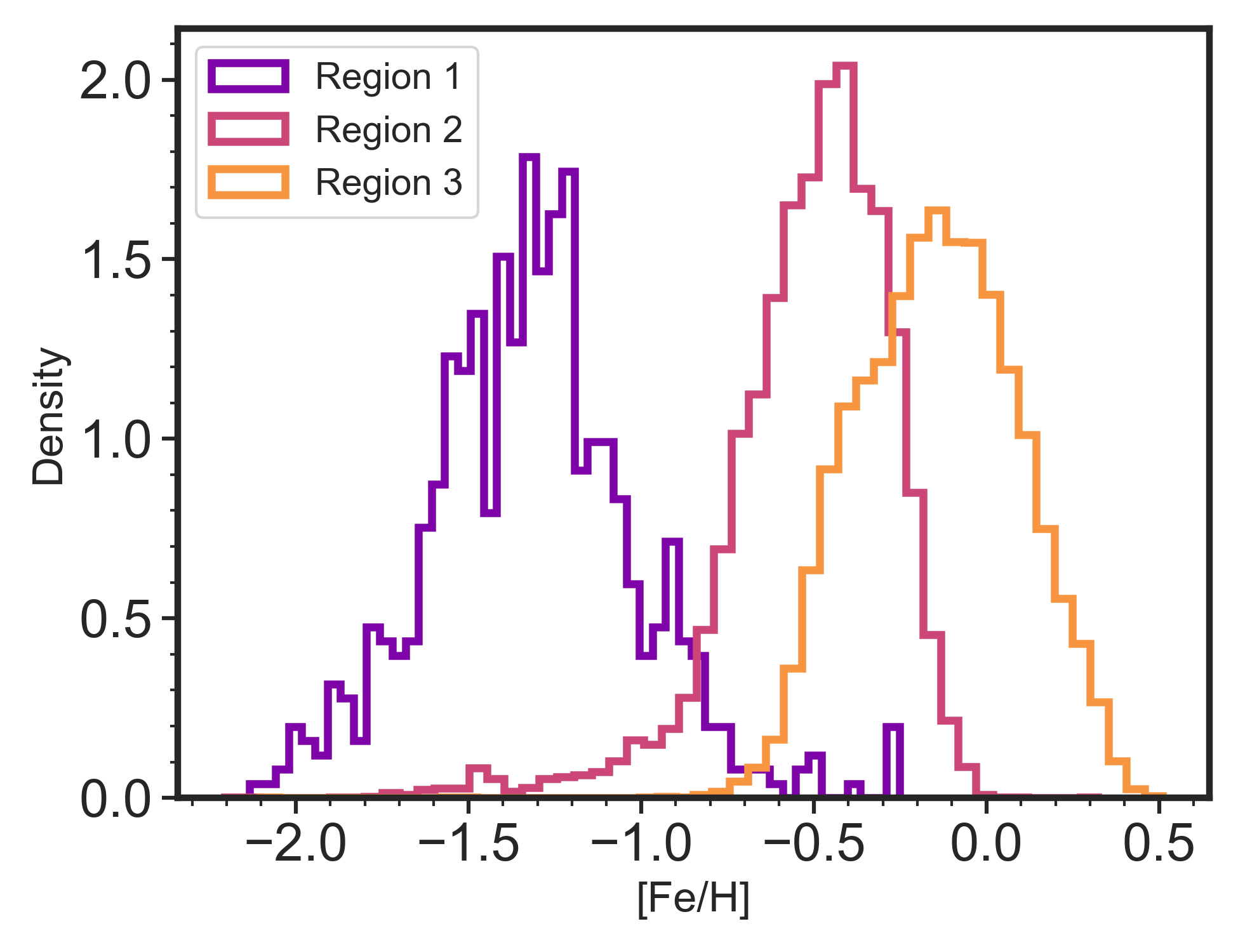}
 \caption{ Metallicity distribution functions of the three regions in [Mg/Mn] vs [Al/Fe] (see Figure \ref{fig:mgmnal}) that correspond to the accreted stars (Region 1), thick disc (Region 2), and thin disc (Region 3).}
 \label{fig:mdf}
\end{figure}

\begin{figure*}
 \includegraphics[width=0.95\textwidth]{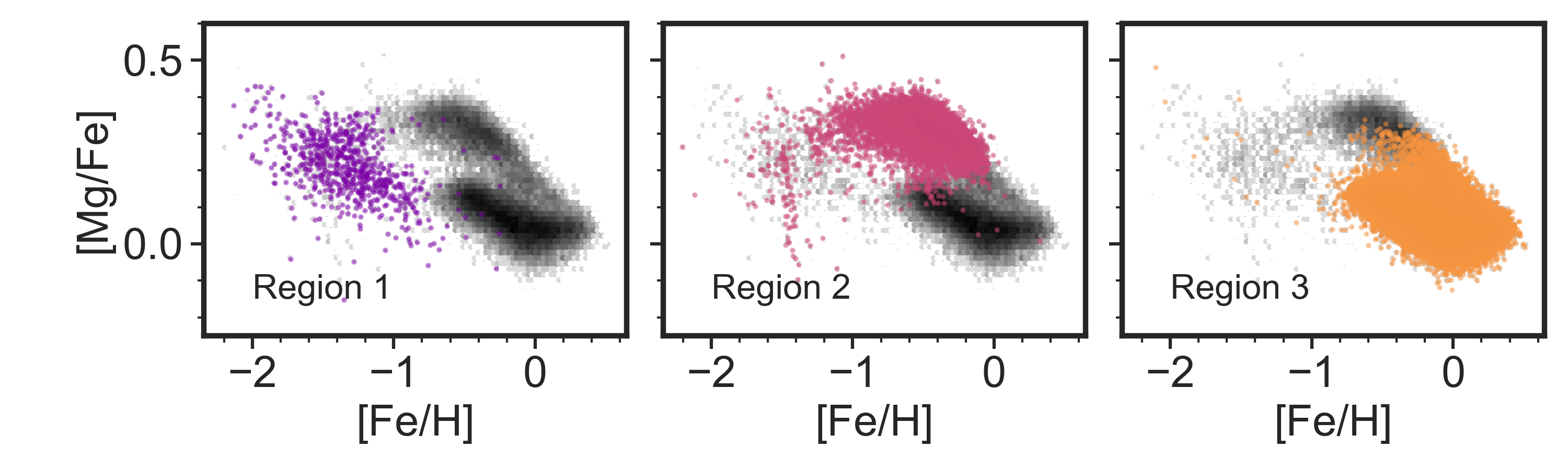}
 \caption{[Mg/Fe] vs [Fe/H] of the three regions in [Mg/Mn] vs [Al/Fe] as defined in Figure \ref{fig:mgmnal} that correspond to the accreted stars (Region 1), thick disc (Region 2), and thin disc (Region 3).}
 \label{fig:mgfe_quads}
\end{figure*}

\begin{figure*}
 \includegraphics[width=\textwidth]{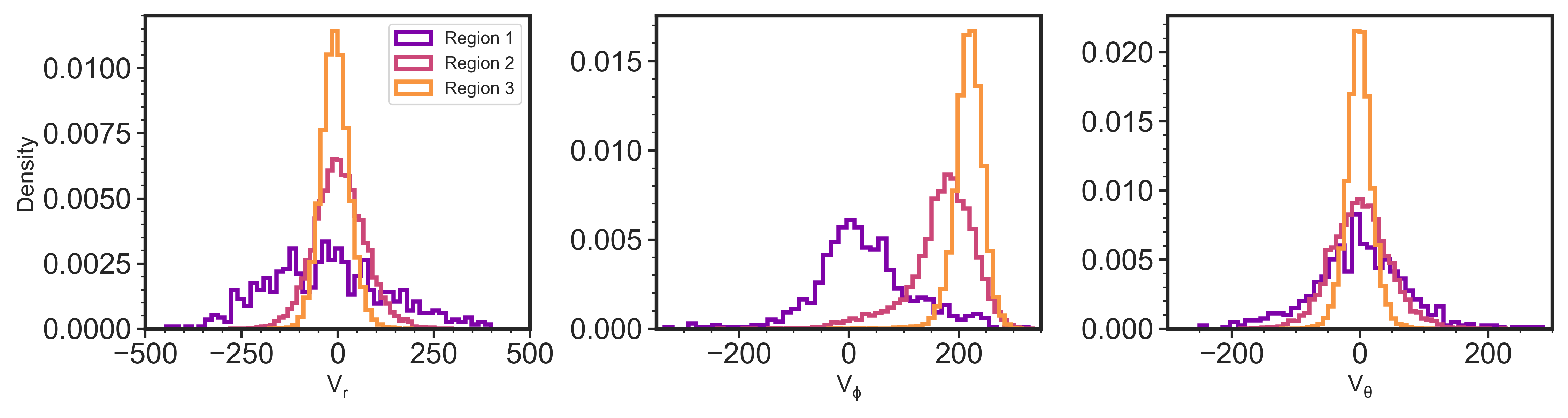}
 \caption{Distributions of kinematics in Galactocentric spherical coordinates showing the radial (\Vr), azimuthal (\Vphi), and polar (\Vtheta)~velocity components of the three regions in [Mg/Mn] vs [Al/Fe] (see Figure \ref{fig:mgmnal}) that correspond to the accreted stars (Region 1), thick disc (Region 2), and thin disc (Region 3). Positive \Vphi~corresponds to rotation with the disc. }
 \label{fig:vels_3reg}
\end{figure*}

The [Mg/Mn] vs [Al/Fe] plane with APOGEE data has extensively been used to select the accreted population of the Milky Way. This is an effective way to distinguish accreted vs in-situ material because Mg is an indicator of core-collapse supernovae, Mn is an indicator of Type Ia supernovae, and Al has been shown to be lower for dwarf galaxies compared to in-situ stars (e.g., \citealt{hasselquist21}). The accreted stars separate nicely as a concentration of stars at [Mg/Mn]$>$0.25 and [Al/Fe] < 0, away from the bulk of the in-situ material as shown in the top panel of Figure \ref{fig:mgmnal}. It is especially effective in selecting GES stars in the Solar neighborhood and inner halo, showing agreement with previous studies in terms of their ages and orbital properties \citep{das20,bonaca20}, as well as in their detailed chemical abundances, especially in the neutron-capture elements \citep{matsuno20,aguado20,carrillo22a}. 

To illustrate how well this plane is able to separate different stellar populations in the Milky Way, we show the diagram split into three different regions as shown in the bottom panel of Figure \ref{fig:mgmnal}. Region 1 (purple) represents the accreted population, region 2 (pink) the thick disc, and region 3 (orange) the thin disc. 

Figure \ref{fig:mdf} illustrates the MDF and Figure \ref{fig:mgfe_quads} shows the [Mg/Fe] vs [Fe/H] for these three regions. The Region 1 MDF, which is associated with the accreted halo population, has a peak at the lowest metallicity of all three regions at [Fe/H] $\sim$ -1.2. This is in line with the typical [Fe/H] peak for GES from previous works \citep{das20,feuillet21,buder22}. There is a smaller secondary peak at [Fe/H] $\sim$ -0.2 which is due to some contribution at lower [Mg/Mn] values, at the boundary with the pink region that corresponds to the thin disc. As shown in Figure \ref{fig:mgfe_quads}, these stars follow the lower track in the [Mg/Fe] vs [Fe/H] plane compared to the higher-[Mg/Fe] in-situ population. At [Fe/H] $<$ -1.5 however, the track changes slope, similar to what has been previously found for GES in the APOGEE data (e.g., \citealt{myeong22}). Recently, \citet{feltzing23} looked into this diagram in greater detail and found that though the majority of stars in this region are accreted, some stars show kinematics more akin to the Milky Way disc. 
Accreted material from other systems can and do exist in this region (see \citealt{horta22} for a detailed exploration of the halo substructures in APOGEE). Nonetheless, the majority of our sample in this region comes from GES given that it dominates the accreted population for $R_{GC}<20$kpc \citep{naidu20} and we are probing a local region given our cuts in the observations. 

The Region 2 MDF has a peak at [Fe/H] $\sim$ -0.4, in line with values for the thick disc \citep{kordopatis13,gaia18}. There is a long tail towards lower metallicity, which is due to contamination from both in-situ halo \citep{gallart19} and accreted halo populations.
The middle panel of Figure \ref{fig:mgfe_quads} shows that the Region 2 stars mainly track the [Mg/Fe] vs [Fe/H] trend for the thick disc, but also show some contamination from stars that are associated with the accreted halo. 

Lastly, the Region 3 MDF has a peak at solar metallicity, much like what is expected for the thin disc of the Galaxy. The boundary between the thin and thick discs are not as distinct in the [Mg/Mn] vs [Al/Fe] plane compared to the boundary of the in-situ population and that of the accreted population (see top panel of Figure \ref{fig:mgmnal}). Thus, there is a bump---not just a tail---for the lower metallicity half of the Region 3 MDF due to contributions from the thick disc. This contamination is especially apparent in the [Mg/Fe] vs [Fe/H] plane showing a considerable portion of the thick disc track being occupied by Region 3 stars in Figure \ref{fig:mgfe_quads}. Nonetheless, based on the MDF and the [Mg/Fe] vs [Fe/H], splitting the different concentrations in the [Mg/Mn] vs [Al/Fe] is a robust way of distinguishing accreted vs in-situ populations from each other. 

In addition to the chemistry, the kinematics for these three regions in the [Mg/Mn] vs [Al/Fe] plane are consistent with the accreted halo, thick disc, and thin disc stellar populations for regions 1, 2, and 3, respectively. In Figure \ref{fig:vels_3reg}, we show the Galactocentric spherical velocity components for these regions. These are defined as follows: Vr is the radial component, $\rm V_{\phi}$ is the azimuthal component, and  $\rm V_{ \theta}$ is the polar component. We derived these quantities following Section 2 in \citet{bird19} and folded in the errors from the distance, proper motions, and radial velocities from Gaia through resampling 100 times and drawing a new velocity each time. The distributions in Figure \ref{fig:vels_3reg} are created from the median spherical velocity component for each star. The uncertainties in the three velocities are $\sim$15 km/s for Region 1, $\sim$5 km/s for Region 2, $\sim$3 km/s for Region 3. 

Region 3, which we associate with the thin disc, shows the lowest dispersion in Vr, $\rm V_{\phi}$, and $\rm V_{ \theta}$ with $\rm V_{\phi}$ $\approx$ 220 km $\rm s^{-1}$ therefore showing rotation with the disc, and Vr and $\rm V_{ \theta}$ centered at zero. 
The region associated with the thick disc (Region 2) shows a larger dispersion in all three velocity components compared to the thin disc, as is expected for a more kinematically hot component. 
In addition, the $V_{\phi}$ for the thick disc is lower than the thin disc's at $\sim$ 170 km $\rm s^{-1}$. Lastly, the accreted halo component (Region 1) shows the largest dispersion in all three spherical velocity components and show no rotation with $V_{\phi}$ centered at 0 km $\rm s^{-1}$. 

Having shown the discerning power of this combination of elements, we therefore use the [Mg/Mn] vs [Al/Fe] to chemically select accreted halo stars in the APOGEE-Gaia crossmatch without any additional cuts in kinematics. To do this less arbitrarily, we broke down the different regions using the Gaussian Mixture Model (GMM) from \texttt{scikit-learn}. We use Bayesian Information Criterion (BIC) to determine the best number of components to describe our data. We ran GMM using 3 to 20 components, and using BIC, we determine the best number of components to be 11. This appears to be too many components to explain the different stellar populations in this chemical plane, with the majority of them in the in-situ region. However, for our purposes, this is less important as we are focusing on the accreted halo population. In fact, there is only one associated component in the accreted halo region as the concentration of these stars is easily separated from the rest as shown in the top panel of Figure \ref{fig:mgmnal}. We selected the stars with $\ge$70\% probability of belonging to the accreted component in the [Mg/Mn] vs [Al/Fe] diagram and consider this the purely chemically-selected GES sample. \citet{das20} and \cite{carrillo22a} make a further cut, i.e., [Fe/H] $\le$ -0.7 to remove contamination from kinematic thick disc stars. However, we do not perform this as we want to compare the different levels and sources of contamination among the various ways GES is selected in their very simplest forms. \textit{With this selection, we have a sample of 356 GES stars.}

\subsection{Dynamical selection of accreted stars in observations}
\label{sec:kinematic_obs}


The more common way for selecting GES stars is through their phase-space information and dynamics. We apply such selections in the APOGEE-Gaia crossmatch, specifically in energy-angular momentum (E-Lz, \citealt{helmi18}; \citealt{horta22}), eccentricity (\textit{e}, \citealt{naidu20}; \citealt{myeong22}), radial action-angular momentum (Jr-Lz, \citealt{feuillet21}; \citealt{buder22}; \citealt{limberg22}), and action diamond (\citealt{myeong19}; \citealt{lane22}) spaces. We note that we \textit{do not} make any other additional cuts, such as in chemistry, in creating these GES samples.  We also explored the effects of observational uncertainties on these dynamical selections. Due to computational time, we folded in the uncertainties from the distance, proper motions, and radial velocities for only 0.5\% of the parent sample to understand their effects. For each star, we drew a new orbital property 100 times given the input parameters and their uncertainties, and inspected the median and standard deviation from this exercise. With this subsample, we note (1) the median difference in the orbital property between the run with and without resampling and (2) the median error from the resampling as follows: energy = (192, 2325) $\rm km^{2}/s^{2}$, Lz = (6, 78) kpc km/s, Jr = (6, 50) kpc km/s, Jz = (2, 22) kpc km/s, eccentricity = (0.01, 0.031), apocentre = (0.07, 0.72) kpc, pericentre = (0.03, 0.22) kpc, and zmax = (0.11, 0.92) kpc. From this test, we can see that the differences and errors in the orbital properties would have a negligible effect on the dynamical selections  because they are significantly smaller in magnitude than the cuts applied.

For the E-Lz method, we applied the same cut in Lz as originally used in \citet{helmi18} i.e., $-1500 < L_{z} < 150$ kpc km/s. Note that the positive (and therefore prograde) part of this selection does not cover as large of a range in Lz but this was originally done to avoid in-situ material. We modify the bound in energy from $-1.8 \times 10^{5}$ $\rm km^{2}/s^{2}$ to $-1.6 \times 10^{5}$ $\rm km^{2}/s^{2}$. This cut was determined from visual inspection to also avoid the region largely dominated by in-situ material and that associated with Heracles in the deeper part of the potential \citep{horta20b}. The difference in the appropriate energy bound is likely due to the difference in the Milky Way potential that we used. \textit{With the E-Lz selection, we have a sample of 630 GES stars. } Similar to our chemical selection, we do not employ additional cuts outside of E and Lz in order to compare these selection methods in a more straightforward way. 

Next, we explore the GES selection in eccentricity. From the APOGEE-Gaia crossmatch, we apply a cut in eccentricity using $e~>0.7$ as similarly done by \citet{naidu20} with the H3 survey. This selection reproduces the highly radial, ``sausage"-like component of the halo in \Vphi~vs \Vr\ plane, which is one of the initial ways GES was discovered with Gaia data \citep{belokurov18}. A massive satellite like the GES would have been affected significantly by dynamical friction, and the orbits of its most bound stars would have been more radialised with time \citep{amorisco17,vasiliev22};  GES, therefore, has a large portion of stars at highly eccentric orbits. As this selection is quite simplistic, it is (1) prone to contamination of high-eccentricity, in-situ stars and (2) misses the low-eccentricity tail of GES.
We keep these caveats in mind throughout our comparisons and again, similar to the two other previously mentioned selections, we do not make any further cuts outside of eccentricity. \textit{With the eccentricity selection, we have a sample of 1,279 GES stars.} 

We also use the radial action, Jr, and angular momentum, Lz of the sample to select highly radial stars in the APOGEE-Gaia crossmatch. This selection, made in the $\sqrt(J_{R})$-Lz space, was originally introduced by \citet{feuillet20} using the SkyMapper and Gaia surveys as a purer selection for GES stars determined from its MDF. The cut is made with 30 $\le \sqrt(J_{R})$ $\le$ 50 $\rm (kpc~km~s^{-1})^{1/2}$ and -500 $\le$ Lz $\le$ 500 kpc km/s. In the original paper, this cut produces a sample of stars with a narrow MDF (e.g., dispersion of 0.34 dex) centered at [Fe/H] = -1.17. 
\textit{With the same Jr-Lz selection, we have a sample of 144 GES stars.} Similar to the other selection methods we have already introduced, we do not make any further cuts in chemistry or kinematics in order to understand the effects of purely selecting in this frame. 

Lastly, we use the action diamond to select highly radial stars. One axis of the action diamond is $(J_{z}-J_{R})/J_{tot}$ and the other is $J_{\phi}/J_{tot}$ where $J_{tot} = |J_{z}| + |J_{R}| + |J_{\phi}|$ and $J_{\phi} = L_{z}$. Along the $(J_{z}-J_{R})/J_{tot}$ axis, values closer to -1 correspond to stars with radial orbits while those closer to 1 correspond to stars with polar orbits. On the other hand, along the $J_{\phi}/J_{tot}$ axis, values closer to -1 correspond to stars with retrograde orbits while those closer to 1 are stars with prograde orbits. We employ a similar selection as \citet{lane22} i.e., $|L_{z}/J_{tot}|<0.07$ and $(J_{z}-J_{R})/J_{tot}<-0.3$ for the GES stars, as they found that this action diamond selection gives the purest sample of GES stars among six other kinematic methods. \textit{With this action diamond selection, we have a sample of 160 GES stars}.


\section{Chemical vs kinematic selection in observations}
\label{sec:chemical_vs_kinematic_obs}

\begin{figure*}
 \includegraphics[width=0.19\textwidth]{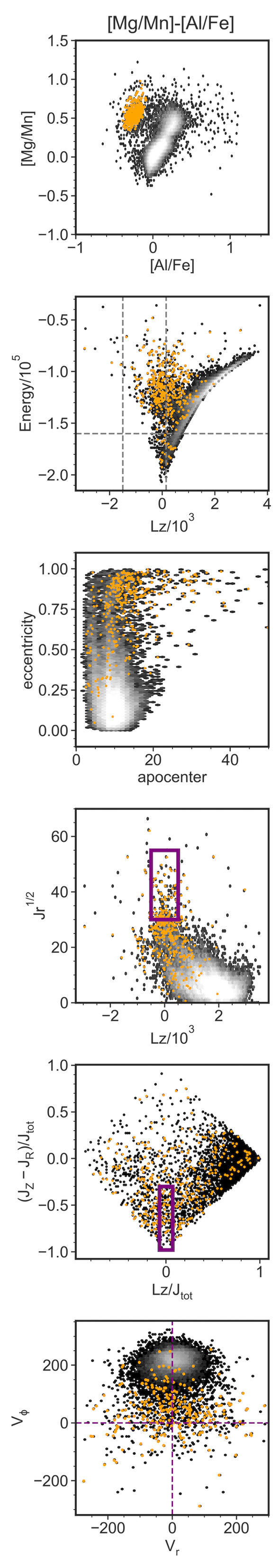}
  \includegraphics[width=0.19\textwidth]{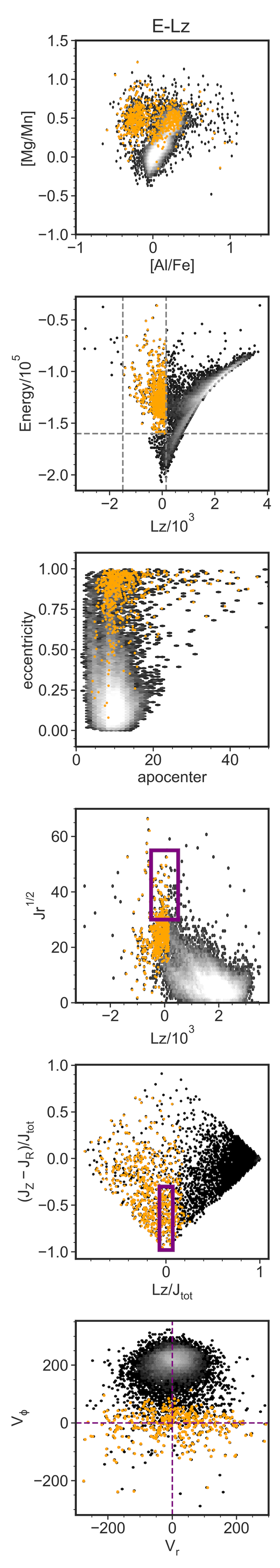}
  \includegraphics[width=0.19\textwidth]{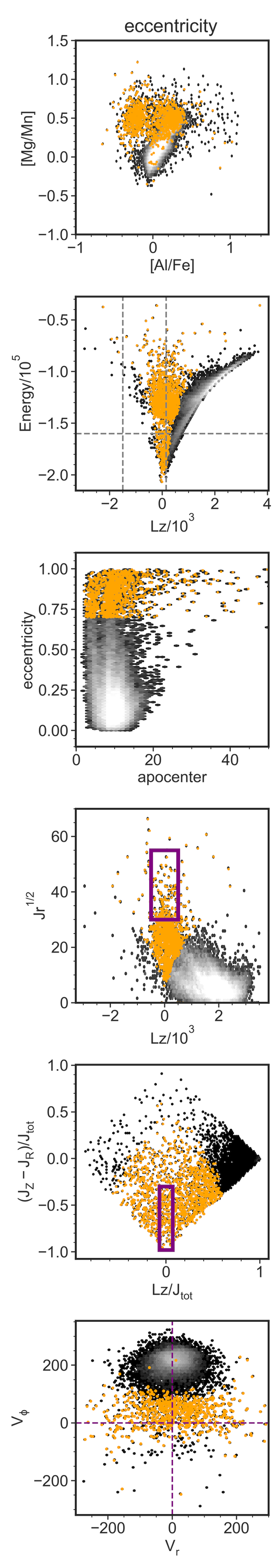}
 \includegraphics[width=0.19\textwidth]{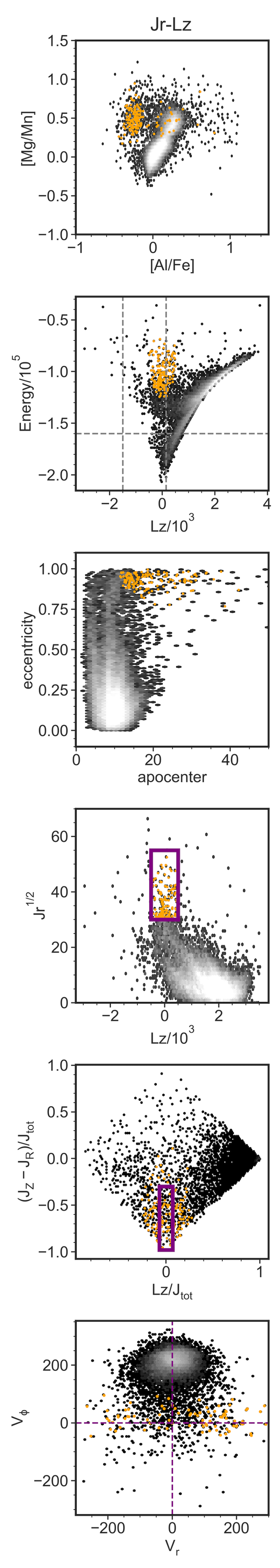}
 \includegraphics[width=0.19\textwidth]{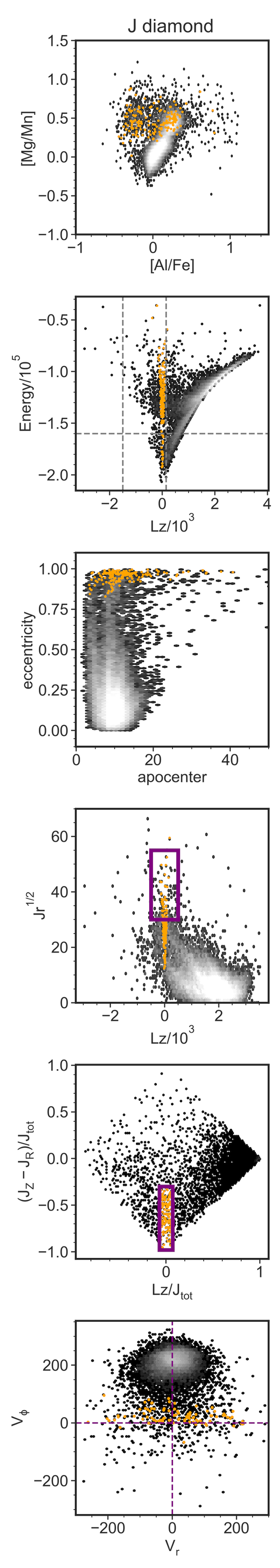}
 \caption{\textit{Different GES samples projected onto the various chemo(dynamical) diagrams.} Columns from left to right: selections in [Mg/Mn]-[Al/Fe], E-Lz, eccentricity, Jr-Lz, and action diamond, respectively. Rows from top to bottom: projections of each GES sample onto [Mg/Mn]-[Al/Fe], E-Lz, eccentricity, Jr-Lz, action diamond, and \Vphi-\Vr. The GES samples are shown in orange and the greyed out background is the parent Gaia-APOGEE sample. The selection box for the Jr-Lz and action diamond methods are marked with a purple box in rows 4 and 5, respectively.} 
 \label{fig:ges_chem_4panels}
\end{figure*}

\begin{figure*}
\centering
 \includegraphics[width=0.98\textwidth]{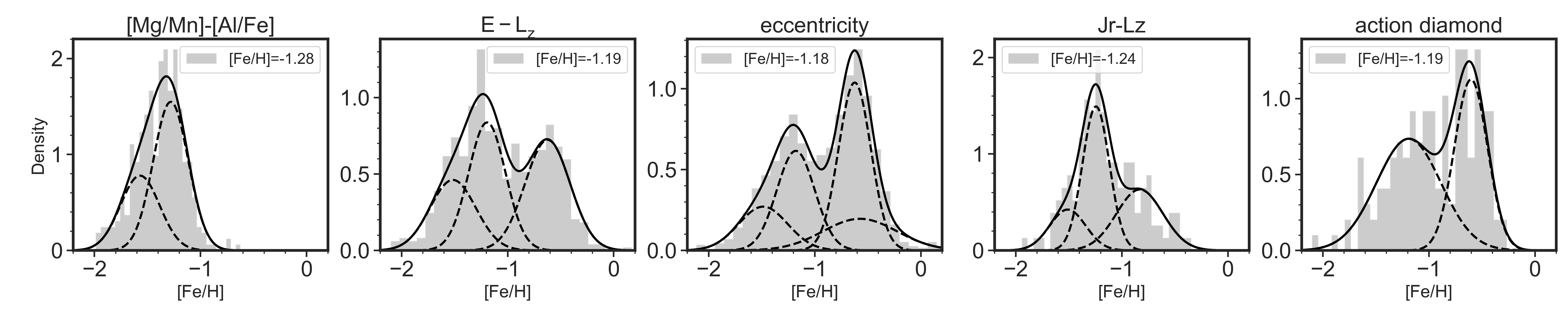}
 \caption{Metallicity distribution function of GES from the different selection methods as indicated in the title of each panel. The black line indicates the total best-fit from the Gaussian mixture modeling while the solid dashed lines are the individual Gaussians. The associated GES [Fe/H] peak is noted in the legend of each subpanel.}
 \label{fig:MDFs_4selection}
\end{figure*}


\subsection{GES samples from the different methods}
We now look at the performance of each selection method by showing the selected GES stars in the other spaces we considered for selection. This is seen in Figure \ref{fig:ges_chem_4panels}---from left to right shows the GES stars (orange markers) selected from [Mg/Mn]-[Al/Fe], E-Lz, eccentricity, Jr-Lz, and action diamond, respectively, projected onto the other selection spaces from top to bottom. We also include \Vphi-\Vr, even though we do not use this space for our selection, because this shows the elongated feature along \Vr~that is attributed to GES. The whole Gaia-APOGEE data set from which we culled out the different GES samples is shown as the greyed out background. In rows 4 and 5 of Figure \ref{fig:ges_chem_4panels}, we mark the selection box for the Jr-Lz and action diamond methods, respectively, with a purple box. 

\subsubsection{[Mg/Mn]-[Al/Fe]} 
\label{sec:obs_gse_chem}
This chemical selection by construction only occupies the accreted area (i.e., Region 1 from previous discussion) in the [Mg/Mn] vs [Al/Fe] diagram. This GES sample is largely at Lz$\approx$0 kpc km/s and with energy $>$ -1.6$\times 10^{5}$ $\rm km^2/s^{2}$. Some GES stars go deeper into the potential with even lower energies, and some stars show retrograde motion, reminiscent of another accreted population, Sequoia \citep{myeong19}. Around 66\% (236) of these GES stars have high eccentricities i.e., $e>0.75$, and this sample has a median apocentre of 13.9 kpc but even reaches values $>$50 kpc. The lower eccentricity stars ($e<0.75$) within this selection have a lower median apocentre of 9.5 kpc. 
We also project these chemically-selected GES stars onto the Jr-Lz space. We have marked the Jr-Lz box as defined in \citet{feuillet21} to guide the eye. The majority of the stars have high Jr, and interestingly, the floor for our sample lies at a lower $\rm \sqrt{Jz}$ bound at $\sim$22 $\rm (kpc~km/s)^{1/2}$ instead of 30 $\rm (kpc~km/s)^{1/2}$ as defined for the box. This could be due to the difference in the potential the stars were integrated in i.e., we use \citet{cautun20} while \citet{feuillet20,feuillet21} use MWPotential14 as included in \texttt{galpy}. However, \citet{buder22} similarly find that the their chemical selection reach lower $\rm \sqrt{Jz}$ than the \citet{feuillet20} selection. There are 129 stars (36\%) below $\sim$22 $\rm (kpc~km/s)^{1/2}$ and these lower Jr stars show higher dispersion in Lz compared to the higher Jr stars. We show these chemically-selected GES stars in the action diamond where we have also marked the GES selection box as used by \citet{lane22} and \citet{myeong19}. This GES selection occupies a large region of this diagram but for the most part have $(J_{z}-J_{R})/J_{tot}<0$ and therefore move more radially. Only 35 stars within the chemical selection overlap with action diamond box. Lastly, the \Vphi-\Vr~of these GES stars are indeed elongated, showing stars with high dispersion in \Vr~at 112 km/s but also a high dispersion in \Vphi~at 78 km/s. A handful of stars (i.e., 54 stars, 15\%) seem to show kinematics more akin to the in-situ population with \Vphi> 100 km/s and \Vr~centered at 0 km/s. This is in line with \cite{feltzing23}, that found Milky Way disc stars overlapping with this GES selection.

\subsubsection{E-Lz} 
\label{sec:obs_gse_elz}

These GES stars occupy a larger space in the [Mg/Mn]-[Al/Fe] plane, covering both the accreted and thick disc regions. Indeed, 309 stars (49\%) are in the region associated with the in-situ material in this chemical space. This sample of GES stars are more retrograde by construction as shown in the E-Lz diagram. Most of the stars are concentrated towards Lz$\sim$0 kpc km/s but there are also regions of stars with more retrograde motions and at higher energies. The majority of this GES sample (82\%) have eccentricities greater than 0.75, showing an increasing range in apocentre with increasing eccentricity. 
In the Jr-Lz space, these stars have a larger range in Jr and extends to the downward retrograde branch that \citet{feuillet21} associates with Sequoia. Similar to the chemical selection, the majority of the stars do not lie within the Jr-Lz selection box. We note, however, that the Jr-Lz box was constructed to select more purely and is therefore a tighter constraint. In the action diamond space, it is clear that the majority of these stars move more radially with $(J_{z}-J_{R})/J_{tot}<0$. This GES selection spans a large range in $(J_{z}-J_{R})/J_{tot}$ and $L_{z}/J_{tot}$, with 141 stars overlapping with the action diamond selection box \citep{lane22}, which is 88\% of the GES stars selected from the action diamond method. These GES stars also show the classic ``Sausage" feature in \Vphi-\Vr~with a noticeably smaller dispersion compared to the [Mg/Mn]-[Al/Fe] selection in the first column.

\subsubsection{Eccentricity}
\label{sec:obs_gse_ecc}
Similar to the E-Lz selection, the eccentricity method covers the regions in the [Mg/Mn]-[Al/Fe] diagram corresponding to both the accreted and in-situ populations, with 824 stars (64\%) that overlap with the in-situ population. This sample of GES stars are also centered around Lz$\sim$0 kpc km/s, although this is by construction as a consequence of using the eccentricity. This selection includes stars that are even deeper in the potential, crossing the energy bound that has been put forth as the region corresponding to the Heracles/Kraken merger \citep[][but see also \citealt{lane22}]{horta20b}. 
Within this sample, there are 272 stars in this lower energy (i.e. $<$ $-1.5 \times 10^{5}$ $\rm km^{2}/s^{2}$) substructure. Projecting these stars onto the [Mg/Mn]-[Al/Fe] diagram, only nine of them are in the accreted region and the rest (263 stars) are in the in-situ region. 

By construction, the stars in this sample have eccentricities larger than 0.75. Among these stars however, there are still some substructures; there is a concentration of stars with apocentres at 6 kpc or less (237 stars, 19\%) and are therefore contained to the center of the Galaxy. Interestingly, these stars also span both the accreted (26 stars) and in-situ (237 stars) regions in the [Mg/Mn]-[Al/Fe] diagram, quite reminiscent of the \textit{Aurora} population \citep{belokurov22,myeong22}. 
In the Jr-Lz space for the eccentricity-selected sample, the majority of the stars lie below the selection box. This diagram also shows more clearly the overlap with the in-situ prograde material, which is apparent as a shell starting at $\rm \sqrt{Jz}$ $\sim$ 5 $\rm (kpc~km/s)^{1/2}$ that curves up and to the right to higher Lz values. It is also quite remarkable that although the E-Lz and eccentricity selections seem the most similar of the methods we have explored, the populations they are probing are quite different in Jr-Lz. The eccentricity-selected GES stars barely go down the low Jr and retrograde branch while the E-Lz selected stars do populate this region. In the action diamond space, these stars are dominated by radial action while constrained in Lz, such that they appear to occupy a smaller diamond in the diagram. By construction, this eccentricity cut selects the elongated feature in \Vphi-\Vr, though interestingly, these stars are not centered at $V_{\phi}$ = 0 km/s. This is mostly driven by the in-situ contamination, as previously noted. 

\subsubsection{Jr-Lz}
\label{sec:obs_gse_jrlz}
The Jr-Lz selection has the least number of stars but it successfully selects mostly accreted stars in the [Mg/Mn] vs [Al/Fe] diagram. This dynamical selection results in 116 out of the 144 stars being in the accreted halo region (Region 1) without any additional cuts, with the remaining 28 stars being in the in-situ regions. This is, in fact, very similar to the contamination found by \citet{limberg22} in their exploration of the Jr-Lz selection for GES stars, i.e., at $\sim$18\%. In the E-Lz plane, these stars are the most restricted in Lz, but are still centered around Lz$\sim$0 kpc km/s. They are also at higher energies compared to the other methods with E $>$ $-1.3 \times 10^{5}$ $\rm km^{2}/s^{2}$. These stars exhibit high eccentricities (i.e., $>$0.75) and apocentres greater than 10 kpc, and they show the elongated feature in \Vphi-\Vr. These stars are bound within the selection box in Jr-Lz since they were selected in this plane, and the density of the stars decrease with increasing Jr. As we noted in the other selection methods, the lower bound in Jr for this selection is higher compared to the others. These stars overlap largely with the GES selection in the action diamond space, perhaps unsurprisingly as both methods select stars with high $J_{R}$.  \citet{feuillet21} similarly used APOGEE (DR16) data, applied the same selection method, and found 299 stars associated with GES. Their GES stars occupy high-\Vr~lobes similar to our Jr-Lz GES sample, although this signature in our work is not as strong due to the smaller sample. 

\subsubsection{Action Diamond}
\label{sec:obs_gse_diam}

Finally, we look at the GES stars selected from the action diamond. In the [Mg/Mn] vs [Al/Fe] diagram, 69 stars (43\%) lie in the accreted region in [Mg/Mn] vs [Al/Fe] while the rest i.e., 91 stars, seem more in-situ-like in their chemistry. Contamination from in-situ material in this selection has been similarly found by \citet{feltzing23} in comparing this method to other selections. This GES sample is centered at Lz$\sim$0 kpc km/s by construction and though most of the stars are at high energies, 19 of them are below $-1.5 \times 10^{5}$ $\rm km^{2}/s^{2}$. This action diamond-selected GES sample has stars with highly eccentric orbits with $e > 0.8$, therefore also occupying the ``sausage" region in \Vphi-\Vr. Though this range in eccentricity is quite similar to that of the Jr-Lz selection, the action diamond method also contains stars with lower apocentres (i.e., $<$6 kpc).  Compared to the Jr-Lz selection which was similarly found to select a purer GES sample, the action diamond also includes stars with lower $J_{R}$, delving into the region that overlaps more with the in-situ material. 

\subsection{Metallicity Distribution Functions}
\label{sec:mdfs}

\begin{table*}
\begin{center}
\caption{\textit{[Fe/H] and mass estimates for GES from the different selection methods.} The columns are labeled as follows: (1) Selection method (2) Number of stars in the GES sample from the selection (3) GES [Fe/H] from the Gaussian Mixture Modeling (GMM) (4) Weight associated with the GES component in the GMM given between 0 to 1 where 1 is the highest (5) Stellar mass, \mstar, derived from the mass-metallicity relationship (MZR) in \citet{kirby13} using the [Fe/H] in column 3 (6 \& 7) \mstar~derived from \citet{ma16} MZR at $z=0$ and 2, respectively (8) \mstar~derived from the observationally-motivated redshift-evolved MZR from \citet{naidu22}, essentially shifting the $z=0$ \citet{kirby13} MZR down by $\sim$0.3 dex (9) halo mass, $\rm M_{halo}$, of GES using the stellar mass-halo mass relations from \citet{behroozi19} at $z=0$ and 2, with the \mstar~from column 5 as input.}

\begin{tabular}{cccccccccc}
\hline \hline
 (1) & (2) & (3) & (4) & (5) & (6) & (7) & (8) & (9) \\ 
method & N & [Fe/H] & GMM & $M_{\star, z=0}$, K13 & $M_{\star, z=0}$, M15 & $M_{\star, z=2}$, M15 & $M_{\star,\rm z-evol}$, N22 & $M_{\rm halo,z=0,2}$, B19 \\  
 &  &  & weight & $10^8$ $M_{\odot}$ & $10^8$ $M_{\odot}$ & $10^8$ $M_{\odot}$ & $10^8$ $M_{\odot}$ & $10^{11}$ $M_{\odot}$ \\
\hline
[Mg/Mn]-[Al/Fe] & 356 & -1.28$\pm$0.03 & 0.64 & 2.37$\pm$1.74 & 1.72$\pm$0.32 & 19.15$\pm$3.24 & 24.51$\pm$20.27 & 0.69, 1.15 \\
E-Lz & 630 & -1.19$\pm$0.03 & 0.37 & 4.53$\pm$1.92 & 2.72$\pm$0.43 & 31.19$\pm$5.18 & 43.66$\pm$26.56 & 0.96, 1.61 \\
eccentricity & 1279 & -1.18$\pm$0.03 & 0.28 & 5.26$\pm$2.96 & 3.02$\pm$0.50 & 34.72$\pm$5.62 & 52.96$\pm$51.17 & 1.04, 1.74 \\
Jr-Lz & 144 & -1.24$\pm$0.02 & 0.47 & 3.13$\pm$2.02 & 2.11$\pm$0.34 & 23.80$\pm$3.73 & 33.90$\pm$23.59 & 0.80, 1.33 \\
action diamond & 160 & -1.19$\pm$0.09 & 0.56 & 4.75$\pm$2.91 & 2.79$\pm$0.47 & 32.37$\pm$5.52 & 46.57$\pm$37.96 & 0.99, 1.65 \\

\hline
\hline
\end{tabular}
\label{tab:MDF}
\end{center}
\end{table*}

We next explore the MDF of the resulting GES samples from each selection method, and these are shown in Figure \ref{fig:MDFs_4selection}. As discussed in the previous section, each method has varying levels of contamination from non-GES stellar populations, both in-situ and accreted (i.e, Sequoia). The contamination is also more prominent when looking at the MDFs, especially for the E-Lz, eccentricity, and action diamond selections that clearly show multiple peaks.

We next broke down each MDF into a mixture of Gaussian distributions to further ascertain the level of contamination and to find the associated distribution and the [Fe/H] peak of GES. We first determined the best number of components by minimizing the Bayesian Information Criterion (BIC) and performed Gaussian Mixture Modeling (GMM) using the \texttt{sklearn} package. The individual Gaussian distributions are shown in Figure \ref{fig:MDFs_4selection} as the dashed lines with the total, best-fit mixture model shown as the solid lines. Now with the different peaks identified, we assigned the GES peak informed by previous works with ranges between -1.4 and -1.1 dex   \citep[e.g.][]{naidu20,feuillet20,das20,bonifacio21,buder22} and list these values in Table \ref{tab:MDF}. Specifically, the MDF peak for the GES are at -1.28, -1.19, -1.18, -1.24, and -1.19 for the chemical, E-Lz, eccentricity, Jr-Lz, and action diamond selections, respectively. It is interesting how the chemical selection provides the lowest peak in the MDF. This difference between the chemical and dynamical selections has indeed been previously noted for the APOGEE data (see discussion in \citealt{buder22}) and is likely a result of such selections methods probing different stellar populations of GES.

We make sense of the other sources of contamination by investigating the origin of the individual metallicity distributions. The negatively skewed MDF from the chemical selection (first panel) necessitates two Gaussians to describe the data best. On one hand, it is known that massive Milky Way satellites have negatively skewed MDFs \citep{kirby13} and in that case, the lower-[Fe/H] peak is most likely associated with the same system i.e., GES. On the other hand, we do find stars that are typically associated with Sequoia on the basis of their dynamics in this GES selection, and in that case, they would be contributing to the lower metallicity tail of the distribution. The lower-metallicity Gaussian has [Fe/H] at -1.57 dex, and is weighted at 36\% from the mixture modeling. Interestingly, Sequoia is also seen to peak at [Fe/H] $\approx$-1.6 \citep{myeong19,naidu20}. However, from Figure \ref{fig:ges_chem_4panels}, the highly retrograde Sequoia component contributes minimally to this selection. Therefore, even if Sequoia is part of the negative metallicity tail, the majority of the tail is likely part of GES. 

A similar lower-[Fe/H] peak is also observed for both the E-Lz, eccentricity, and Jr-Lz selections. These stars are largely associated with the GES based on their dynamics. 
The E-Lz, eccentricity, and action diamond selections all have clear higher-[Fe/H] peaks corresponding to the in-situ material. The prominence of the higher [Fe/H] peak roughly tracks the contamination measured for the E-Lz and eccentricity selections: their weights from the GMM are 39\% and 55\%\footnote{Here, we combine the weights from the two Gaussians at [Fe/H] > -1}, respectively while the in-situ contamination as determined from Sections \ref{sec:obs_gse_elz} and \ref{sec:obs_gse_ecc} are 47\% and 64\%. On the other hand, though the high [Fe/H] Gaussian component in the action diamond selection has higher density in Figure \ref{fig:MDFs_4selection}, the GES component contributes a larger weight at 56\% due to the large dispersion associated with it. Interestingly, the Jr-Lz selection is best fit with three Gaussians. This is surprising as the motivation for this selection comes from producing a well-behaved, normal MDF as determined by \citet{feuillet20}. However, this seems to be due to the spatial cut in z, causing a larger portion of the higher metallicity population to be missed. Lastly, the action diamond method is best fit with two Gaussians, with a higher-metallicity component corresponding to the in-situ material as previously noted in Section \ref{sec:obs_gse_diam} and seen in Figure \ref{fig:ges_chem_4panels}.

On the basis of the number of components, the [Mg/Mn]-[Al/Fe] and the action diamond selections seem the purest in selecting GES stars. These selections also have the highest weights for the Gaussian component associated with GES, at 64\% and 56\%, respectively. The MDFs of massive Milky Way satellites show that the negative skew such as that from the [Mg/Mn]-[Al/Fe] selection \textit{is} what is to be expected for GES. The action diamond has previously been shown to purely select GES stars as well, but we find that this still produces a clear multimodal MDF without any further cuts. Another pure selection is the Jr-Lz method that although needed to be fitted with three Gaussian components, shows significant purity in terms of [Mg/Mn] vs [Al/Fe] (see Section \ref{sec:obs_gse_jrlz}). In addition, this selection weighs the GES as the dominant component at 47\%. For the E-Lz, eccentricity, and action diamond methods that all show multimodal MDFs, the GMM tends to give a higher [Fe/H] for GES. In fact, we find that with increasing contribution from the higher-[Fe/H] Gaussian component---i.e., going from the [Mg/Mn] vs [Al/Fe], to Jr-Lz, to E-Lz, to action diamond, to the eccentricity method---the GES component goes to higher [Fe/H] as well.   

\subsection{Stellar Masses from Mass-Metallicity Relationship}

We now use the [Fe/H] from the GMM of the different selection methods to estimate the stellar mass of GES. We do this by taking advantage of the mass-metallicity relationship (MZR). We use the relation defined from observations of satellites \citep{kirby13} in the local universe, as well as the MZR at $z=0$ and 2 determined from the FIRE simulations which \citet{ma16} showed agrees with observations for $z=0$ \citep{Tremonti04} up to $z\sim3$ \citep{mannucci09}. 
We take the $z=2$ relation to roughly illustrate the MZR at the time that GES was accreted onto the Galaxy. We also calculate the stellar masses from the redshift-evolved MZR as determined from disrupted versus intact satellites \citep{naidu22}---this is essentially shifted by $\sim$0.3 dex from the $z=0$ relation from \citet{kirby13}. These stellar mass estimates are listed in Table \ref{tab:MDF}.

The [Mg/Mn]-[Al/Fe] method gives the lowest [Fe/H] for GES and therefore gives the lowest stellar mass estimate across the different MZRs. Conversely, the eccentricity method has the highest [Fe/H] and resulting stellar mass for GES. From the \citet{kirby13} MZR, the mass for GES has a range of $2.37 - 5.26 \times 10^{8}$ $\rm M_{\odot}$, a factor of $\sim$2.2 difference depending on the selection method. From the \citet{ma16} redshift-evolving MZR, the $z=0$ stellar masses are an order of magnitude smaller than the $z=2$ stellar masses, which has a range of $1.91 - 3.47 \times 10^{9}$ $\rm M_{\odot}$. Using the redshift-evolved MZR determined from \citet{naidu22} gives the highest stellar mass range of $2.45 - 5.30 \times 10^{9}$ $\rm M_{\odot}$. 

With the many ways that the GES stars are defined, there are as many (or even more) ways that its stellar mass has been derived through stellar density \citep{mackereth20}, N-body simulations \citep{naidu21b}, matching to cosmological hydrodynamical simulations \citep{mackereth19}, globular clusters \citep{kruj20,callingham22}, and chemical evolution modeling \citep{fernandezalvar18,vincenzo19,hasselquist21} that give a wide range of total stellar masses spanning an order of magnitude, from $3 \times 10^{8} - 7 \times 10^{9}$ $\rm M_{\odot}$. We similarly find a wide range in stellar masses based on the MZR which is modulated by the accretion redshift. 

Next, we determine the halo mass of GES. We adopted the redshift-evolving stellar mass-halo mass (SMHM) relation from \citet{behroozi19} and calculated the halo mass at $z=0$ and $z=2$ which are listed in Table \ref{tab:MDF}. Because we want to be as observationally motivated as possible, we use the stellar masses derived from the \citet{kirby13} MZR at $z=0$ in calculating the halo mass of GES. The GES halo mass has a range of $0.69 - 1.04 \times 10^{11}$ $\rm M_{\odot}$ with the $z=0$ relation and a range of $1.15 - 1.74 \times 10^{11}$ $\rm M_{\odot}$ with the $z=2$ relation. This puts the GES-Milky Way total mass merger ratio at $11-17$\% using the total Milky Way mass from \citet{deason21} pre-LMC infall and the $z=2$ GES halo mass.  On average however, the Milky Way halo mass at $z=2$ would have been $\sim$20\% of its present day mass, thus increasing the merger ratio to $56-85$\%. On the other hand, the stellar mass merger ratio for the GES-Milky Way collision has a range of $2-5$\% at $z=0$ and $19-53$\% at $z\sim2$ if we similarly assumed as \citet{helmi18} that the stellar mass of the Milky Way was $\sim 10^{10}~\rm M_{\odot}$ at the time of the merger.  In comparison, \citet{grand20} showed that the GES-Milky Way stellar mass merger ratio could be as low as 5\% at infall with the Auriga simulations, while \citet{helmi18} derived a total mass merger ratio of 25\% to produce the Toomre diagram in the observations. This is visibly a wide range of merger ratios for the Milky Way-GES collision, with the redshift evolution of the GES and Milky Way masses being a main factor. With that said, we similarly find the merger ratios are larger for the total mass than the stellar mass, as previous studies have noted.

A huge caveat in deriving the stellar and total mass of GES is the redshift dependence of the MZR and the SMHM, both pushing our estimates to higher values. We therefore aim to be conservative and report the mass estimates from the $z=0$ relation from observations as lower limits to the true GES mass. Based on our exploration of the MDFs (Section \ref{sec:mdfs}), we deem the [Mg/Mn] vs [Al/Fe] method to be the best in selecting GES stars in the observations, with $M_{\star} = 2.37 \pm 1.74 \times 10^{8}~M_{\odot}$ and $ M_{\rm halo} = 0.69 - 1.15 \times 10^{11}~M_{\odot}$. Though this is a lower limit, we can also place an upper limit to the total stellar mass of GES such that it does not exceed the total stellar halo mass i.e., $\sim$1.4 $\times 10^{9} M_{\odot}$ \citep{deason19}.

We have so far explored the different ways we select GES stars in observations, their respective sources and level of contamination, and the resulting stellar and halo mass estimates of the GES progenitor. Depending on the selection method, the stellar mass estimate could differ by a factor of 2, but depending on the adopted redshift, it could differ by a factor of 10. In the next section, we benchmark our different selection methods with simulations where we actually know where the stars come from---i.e., from GES or not.

\section{Accreted stars from a simulation perspective}
\label{sec:simulation_data}

\begin{table*}
\begin{center}
\caption{Properties of Auriga halos with GES-like systems. The columns are labeled as follows: (1) Auriga halo that contains a GES progenitor, (2) GES peak \mstartrue~given in $\rm 10^{9} M_{\odot}$, (3) GES peak total mass $\rm M_{total}$ (i.e., $\rm M_{200}$) given in $\rm 10^{11} M_{\odot}$, (4) Milky Way-like host's peak $\rm M_{\star}$ in $\rm 10^{11} M_{\odot}$, (5) Milky Way-like host's peak $\rm M_{total}$ in $\rm 10^{12} M_{\odot}$, (6) Average [Fe/H] of GES stars, (7) Percentage of GES stars in the observational window as described in Section \ref{sec:simulation_data}. }
\begin{tabular}{ccccccc}
\hline \hline
(1) & (2) & (3) & (4) & (5) & (6) & (7) \\
halo & GES $\rm M_{\star}^{true}$ & GES $\rm M_{total}$ & host $\rm M_{\star}$ & host $\rm M_{total}$ & \feh~ & \% GES in window \\

 & $10^{9}\rm M_{\odot}$ & $10^{11}\rm M_{\odot}$ & $10^{11}\rm M_{\odot}$ & $10^{12}\rm M_{\odot}$ & &  \\
\hline
Au-5 & 3.83 & 1.26 & 0.71 & 1.19 & -0.39 & 12.78 \\
Au-9 & 1.88 & 1.76 & 0.63 & 1.16 & -0.85 & 6.47 \\
Au-10 & 0.97 & 0.39 & 0.62 & 1.02 & -0.65 & 1.29 \\
Au-15 & 2.53 & 1.26 & 0.43 & 1.04 & -0.50 & 8.22 \\
Au-17 & 0.38 & 0.33 & 0.79 & 1.02 & -0.98 & 0.83 \\
Au-18 & 1.44 & 0.75 & 0.84 & 1.39 & -0.70 & 1.62 \\
Au-24 & 2.56 & 1.09 & 0.77 & 1.57 & -0.60 & 4.16 \\
Au-26 & 10.54 & 3.33 & 1.14 & 1.72 & -0.39 & 19.18 \\
Au-27 & 4.08 & 1.72 & 1.03 & 1.85 & -0.53 & 9.40 \\
\hline
\hline
\end{tabular}
\label{tab:auriga_halos}
\end{center}
\end{table*}

In the first part of this work, we investigated the different selections for GES stars in observational data. Now we check which of these methods is best in selecting GES stars using simulations where we have absolute knowledge of which stars belong to the GES-like progenitor. Through this, we hope to answer the question: \textit{Can we really pick and choose?} 

We use the Auriga hydrodynamical simulations \citep{grand17}, which contain 30 high-resolution, cosmological zoom-in simulations of Milky Way-mass haloes (i.e., with virial mass\footnote{Virial mass in the simulations is the mass within a sphere where the mean matter density is 200 times the critical density.} $1-2 \times 10^{12} M_{\odot}$) selected from the dark matter-only $\rm 100^{3} Mpc^{3}$ periodic box of the EAGLE project \citep{schaye15,crain15}. The cosmological parameters were adopted from the Planck Collaboration \citep{planck14}. These selected haloes were resimulated with the \textsc{AREPO} code \citep{springel10} at higher resolution. In this work, we use the resolution level named Level 4 in \citet{grand17} with mass resolution of $\sim 3 \times 10^{5} \rm M_{\odot}$ for the dark matter particles and $\sim 5 \times 10^{4} \rm M_{\odot}$ for the gas cells.

The simulation has a comprehensive prescription for galaxy formation physics that includes primordial and metal-line cooling, star formation and stellar feedback, chemical enrichment (from core-collapse supernovae, Type Ia supernovae, and winds from asymptotic giant branch stars), a sub-grid model for the interstellar medium, black hole formation and feedback, uniform photoionizing UV background, and magnetic fields (see \citealt{grand17} for details). Notably, we use Auriga because it has been shown to contain Milky Way systems with GES-like mergers in the past, as explored in detail by \citet{fattahi19}. We use the \cite{fattahi19} dataset and briefly describe it below, but we refer the reader to the original paper for further details. 

In \citet{fattahi19}, star particles were considered ``in-situ" if they were bound, according to the \textsc{SUBFIND} algorithm \citep{springel01}, to the main progenitor of the Milky Way analogue at their formation time \footnote{In practice this is determined at the snapshot immediately following the birth time.}. If the formation time is at $z>3$, $z=3$ association is adopted. 
Star particles bound to the main halo at $z=0$ but that were previously born in a different halo (in the snapshot after the time of formation) were considered ``accreted" \footnote{From this definition, stars that formed from the gas stripped from the satellite are considered in-situ.}.  

From within the Auriga sample, \citet{fattahi19} determined a subset of ten halos that contain accreted stars exhibiting high orbital anisotropy, $\beta$ $>$0.8 and high metallicity, [Fe/H] $\sim -1$, reminiscent of GES in the observations. They find that the stars contributing to the ``sausage" feature in the simulations come from a single progenitor in most cases, and in fact the most massive progenitor to the halo with stellar mass of $10^9-10^{10}~\rm M_{\odot}$ accreted 6-10 Gyr ago. We use this sub-sample of Auriga halos in exploring the different observational selection methods when applied to the simulations. The properties of these halos are listed in Table \ref{tab:auriga_halos}. We do not include Auriga 22 although it was in the GES sample from \citet{fattahi19} as our observational cuts result in too few star particles for our analysis. 

For the dynamical properties, we use the orbital energy and actions (specifically Lz and Jr), as well as the apocentre and pericentre distances for star particles, determined by \citet{callingham22} for the Auriga halos using AGAMA \citep{vasiliev19}. This is especially important in selecting GES stars in similar ways compared to the observations.

To approximately recreate the same observable area, we limit our sample to $\sim$10 kpc from an arbitrary solar viewpoint at 8.3 kpc and with Galactocentric radii larger than 3 kpc to avoid the bulge region. We similarly avoid the disc and apply a cut in |z| $>$ 1 kpc. The GES stars contribute between $1-20$\% to the stellar population in the observational window and we include these values in Table \ref{tab:auriga_halos}.  With the sample of stars in this region for the nine MW-GES halos, we applied the different GES selections which we discuss in the next section.



\subsection{Selection of accreted stars in simulations}
\label{sec:kinematic_sims}

\begin{figure*}
 \includegraphics[width=0.24\textwidth]{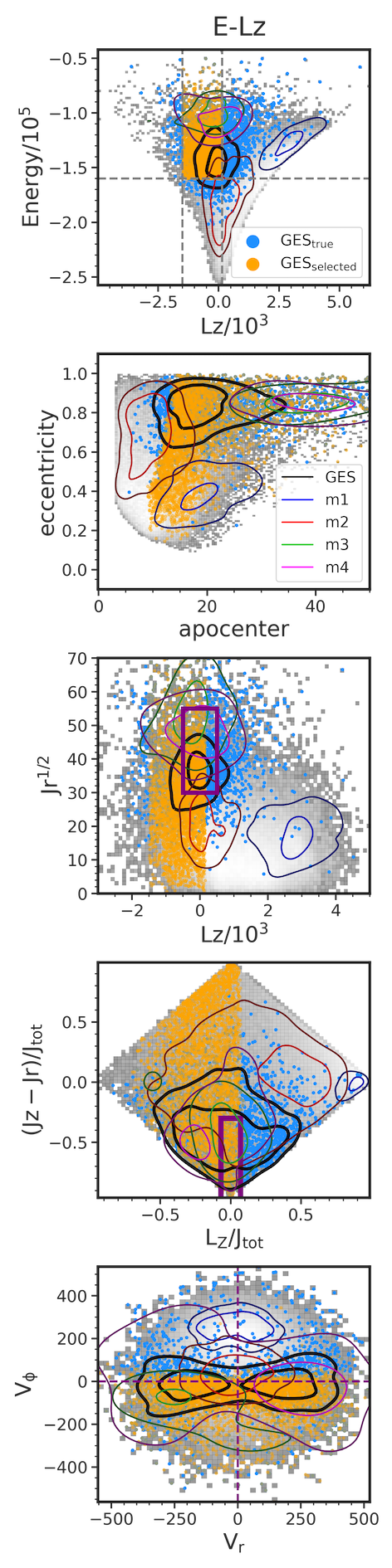}
 \includegraphics[width=0.24\textwidth]{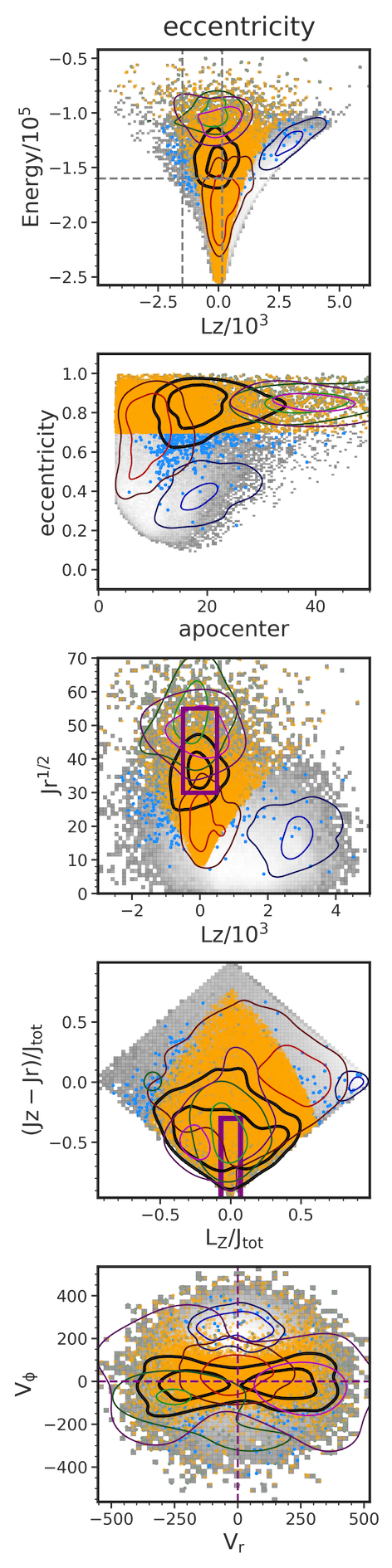}
 \includegraphics[width=0.24\textwidth]{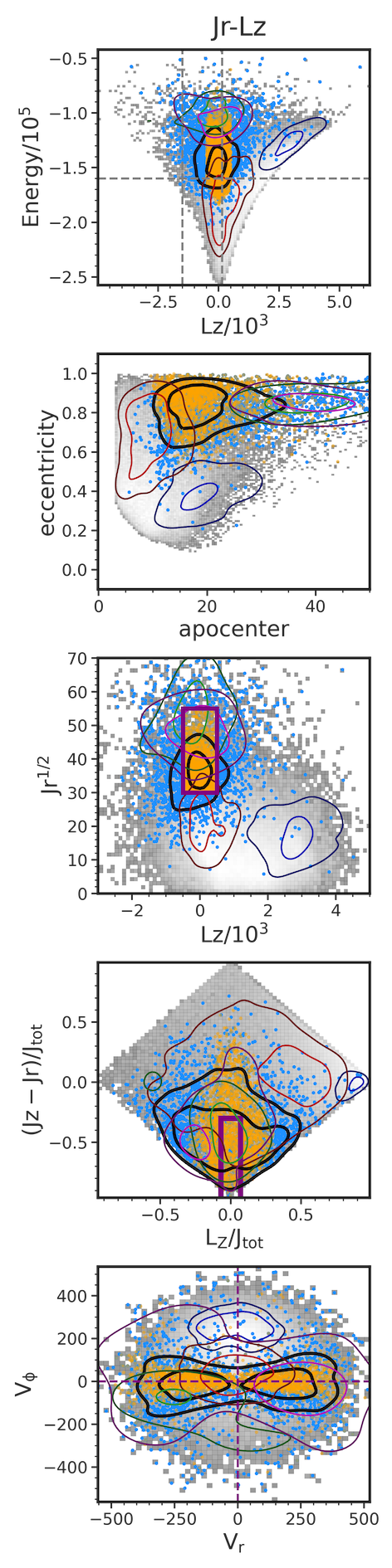}
 \includegraphics[width=0.24\textwidth]{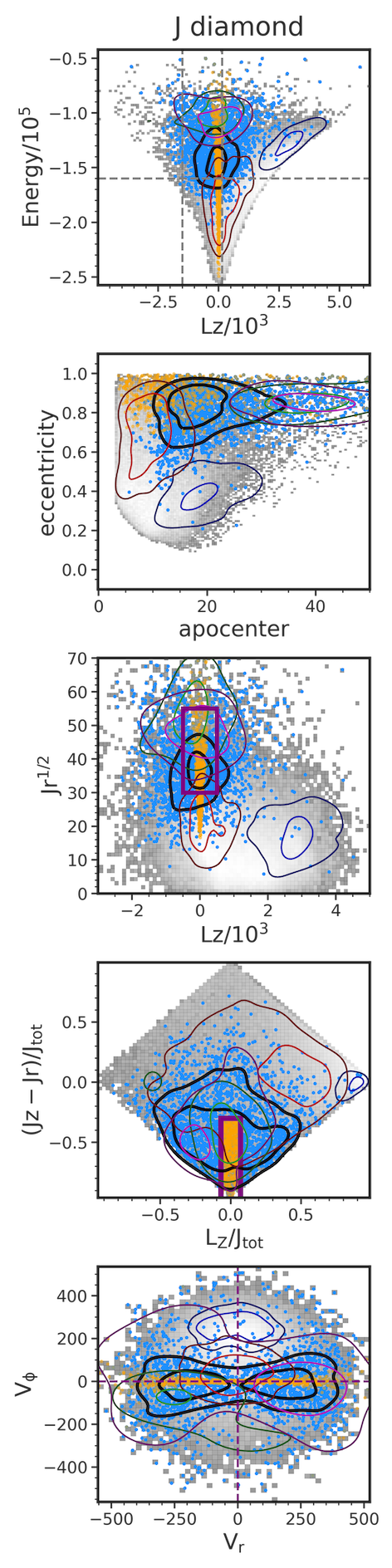}
 \caption{\textit{Different GES samples in Auriga 18 projected onto the various selection diagrams.} Columns from left to right: selections in E-Lz, eccentricity, $\rm \sqrt{Jr}$-Lz, and action diamond, respectively. Rows from top to bottom: projections of each GES sample onto E-Lz, eccentricity, $\rm \sqrt{Jr}$-Lz, action diamond, and \Vphi-\Vr. The \textit{selected} GES stars in the observational window are orange while the \textit{true} GES stars in the same galactic region are light blue. All the stars in the observational window (which is dominated by the in-situ material) is shown as the grey background. 
 The five most massive contributors to the accreted halo in the observable region are shown as contours with  GES-black, m1-blue, m2-red, m3-green, and m4-magenta. }
 \label{fig:ges_panels_sims}
\end{figure*}

\begin{table}
\begin{center}
\caption{Properties of the next four most massive accreted satellites (Column 1) of Au-18 after GES. The rest of the columns are defined as follows: (2) peak \mstartrue~given in $\rm 10^{9} M_{\odot}$, (3) peak $\rm M_{total}$ given in $\rm 10^{11} M_{\odot}$, (4) Average [Fe/H] of stars associated with each progenitor, (5) Percentage of stars in the observational window associated with each progenitor.}
\begin{tabular}{cccccccc}
\hline \hline
halo & \mstar & $\rm M_{total}$ & \feh~ & \% in window \\
 & $10^{9}\rm M_{\odot}$ & $10^{11}\rm M_{\odot}$ & &  \\
\hline
m1 & 0.55 & 0.25 & -0.82 & 0.43 \\
m2 & 0.25 & 0.21 & -0.80 & 0.69\\
m3 & 0.07 & 0.11 & -0.93 & 0.02\\
m4 & 0.06 & 0.09 & -1.27 & 0.03\\
\hline
\hline
\end{tabular}
\label{tab:au18}
\end{center}
\end{table}

\begin{table*}
\begin{center}
\caption{Purity and completeness in terms of percentage (\%) from the different selection methods for the nine Auriga halos investigated in this work.}
\begin{tabular}{@{\extracolsep{8pt}}c|cc|cc|cc|cc}
\hline \hline
halo & \multicolumn{2}{c}{E-Lz} & \multicolumn{2}{c}{Eccentricity} & \multicolumn{2}{c}{Jr-Lz} & \multicolumn{2}{c}{Action diamond} \\
\cline{2-3} \cline{4-5} \cline{6-7} \cline{8-9}
& Purity & Completeness & Purity & Completeness & Purity & Completeness & Purity & Completeness \\
\hline
Au-5 & 45 & 52 & 41 & 78 & 66 & 24 & 56 & 10 \\
Au-9 & 16 & 41 & 15 & 58 & 23 & 13 & 20 & 6 \\
Au-10 & 3 & 39 & 6 & 83 & 12 & 23 & 11 & 10 \\
Au-15 & 14 & 43 & 21 & 52 & 29 & 8 & 23 & 5 \\
Au-17 & 9 & 36 & 5 & 71 & 15 & 23 & 8 & 6 \\
Au-18 & 29 & 49 & 13 & 83 & 33 & 33 & 23 & 11 \\
Au-24 & 27 & 45 & 14 & 62 & 35 & 24 & 21 & 7 \\
Au-26 & 46 & 4 & 38 & 46 & 40 & 10 & 41 & 4 \\
Au-27 & 24 & 3 & 28 & 54 & 33 & 9 & 32 & 5 \\
\hline
\textbf{Median} & \textbf{24} & \textbf{41} & \textbf{15 }& \textbf{62} & \textbf{33} & \textbf{23} & \textbf{23} & \textbf{6}\\ 
\hline \hline
\end{tabular}
\end{center}
\label{tab:pure_comp}
\end{table*}


We highlight Auriga 18 (Au-18) to illustrate the different methods of selecting GES stars but note that we apply these selections to all of the halos in Table \ref{tab:auriga_halos}. Following \citet{callingham22}, we also look at the next four most massive contributors (labelled m1-m4) to the stellar halo of Au-18 listed in Table \ref{tab:au18}, with the most massive progenitor corresponding to GES. This is to give an idea of the other accretion events, in addition to the in-situ material, that can affect the purity and completeness of each selection method. 

We used the same selection in the simulations as in the observations for the E-Lz, eccentricity, Jr-Lz, and action diamond methods (see Section \ref{sec:observation_data}). Unfortunately, the simulations do not contain element abundances for Mn and Al, making the exact comparison with the chemical selection unavailable. However, \citet{tonrud22} have shown that the chemistry in Auriga, specifically [$\alpha$/Fe] and [Fe/H], are very promising in distinguishing accreted versus in-situ stars in the disc using neural network models. We note that the different assembly histories and orbital evolution of these halos (by virtue of being taken from a cosmological simulation) pose differences in where exactly the GES stars lie in these diagrams, as well as where stars from other progenitors lie. However, the selections in observations are generally good approximations for the simulations. In the end, we aim to compare the performance of the different selections with respect to each other within the same halo, so adopting the cuts from the observations is a reasonable approach.

Figure \ref{fig:ges_panels_sims} shows the different GES samples (orange) in the observable window selected in E-Lz, eccentricity, Jr-Lz, and action diamond methods, from left to right. The `true' GES stars in the defined observable window are also shown in blue. The contours indicate the density of stars from the top five massive merger events that contributed to the stellar halo and are assigned as follows: GES-black, m1-blue, m2-red, m3-green, and m4-magenta. The background grey histogram shows the density of all the stars in the observable window, which is dominated by the in-situ stars. Lastly, all these stellar populations are projected onto the E-Lz, eccentricity-apocentre, $\rm \sqrt{Jr}$-Lz, action diamond, and \Vphi-\Vr~diagrams from top to bottom.

It is reassuring that for Au-18, the different selection methods \textit{do} select the GES stars although to varying degrees. For all methods, the majority of the in-situ material in the observable window is avoided by the GES selection i.e., the selected GES stars overlap less with the bright grey regions of the background 2D histogram. This overlap is the greatest for the eccentricity selection, followed by E-Lz, then the action diamond, then lastly by Jr-Lz. We can also compare these GES selections to other accreted material. For example, m3 (green) and m4 (magenta) are both at higher energy and high eccentricity, so there are contaminants from these accretion events in the eccentricity selection without making further cuts in other properties. In the E-Lz selection method, the contamination from m2 (red) is greatly reduced because they are at lower energies, while these stars remain in the GES selection if we only make a cut in eccentricity. The m2 (red), m3 (green), and m4 (magenta) accretion events also overlap with the action diamond box, though we can see in the other selection diagrams that making additional cuts in energy and apocenter will remove these contaminants. The Jr-Lz method seems to be the most selective, having the least overlap with the other accretion events. With that said, the largest contaminant in each GES selection \textit{is} still the in-situ material, which will be discussed further in Section \ref{sec:mdf_sims_selections}. 

Again, we note that the overlap among the GES, in-situ, and other accreted stars is dependent on the halo and its assembly history, and we are merely showing the case for Au-18. However, this does give an idea of what contributes to our calculated purity and completeness values for each selection method. These values are calculated with respect to the GES stars that are within the observable window, instead of the total GES population. We do these for all nine halos, and list these values in Table \ref{tab:pure_comp}. There is variety in the purity and completeness from halo to halo, as some halos are more GES-like than others. However, in general, the eccentricity method performs best in terms of completeness (62\%) and the Jr-Lz method is best in terms of purity (33\%). Conversely, the eccentricity method obtains the least pure sample (15\%) while the action diamond method is the least complete (6\%). 

Now that we have a sample of GES stars from the different methods in the simulations, we can look into the MDFs from these selections vs the real MDF of GES, both for the total population and the observable window. 


\subsection{Metallicity distribution function}

\subsubsection{Total vs observed GES population}
\label{sec:mdf_sim_allvobs}

One way of distinguishing accreted vs in-situ material is by looking at their chemistry. We have shown earlier in the observations that regardless of the selection method, GES has a distinct MDF from the in-situ material centered at lower [Fe/H]. We similarly investigate this in the simulations. In Figure \ref{fig:mdf_sims_gse_ins}, we show the normalized MDF of the corresponding GES in Au-18 as well as the in-situ material both for the total population (top panel) and the population within the observed window (bottom panel). We will refer to the metallicity from the total stellar populations as \fehsimall~while those from the stellar populations in the observational window as \fehsimwindow. We also include a lower-mass accreted system with \mstar~$= 1.4 \times 10^{7} \rm M_{\odot}$ for additional comparison. Lastly, we mark the median [Fe/H] from the total population (dotted line) and those only in the observational region (dot-dash line) for the in-situ stars (purple) and GES (orange). 

For the total population (top panel), there is a distinct progression of higher metallicity for higher stellar mass systems (see also \citealt{fattahi20}). That is, the GES MDF is centered at lower \fehsimall~compared to the in-situ material whilst at higher \fehsimall~compared to the lower mass accreted system. This stellar mass-metallicity relationship in the simulations is encouraging as we can potentially use the metallicity of the selected GES stars to estimate their progenitor's stellar mass as we have done in the observations.

In addition, the shapes of the MDF seem to be quite telling as well. The MDF of GES is negatively skewed, similar though to a lesser degree to that of the in-situ material. On the other hand, the MDF of the lower mass system is wider, and has more of a platykurtic distribution. These are reminiscent of the different MDFs of more massive vs less massive dwarfs around the Milky Way \citep{kirby13}.

The MDFs for the observed window (bottom panel) however are quite different, due to multiple factors. The biggest factor is our spatial cut in z. This preferentially removes the highest metallicity component of the in-situ material i.e., the star-forming disc. Therefore, the GES and in-situ MDFs seem much closer to each other. 
The spatial cut in galactocentric radius ($\lesssim$20 kpc) also preferentially selects the higher metallicity component of the GES population. This is because in the Auriga simulations, GES-like systems have been shown to have a negative metallicity gradient before merging with the Milky Way-like galaxy, wherein its central regions have higher [Fe/H] compared to its outskirts \citep{orkney23}. The more centrally-located stars pre-merger are more bound and stripped later, ending up at lower galactocentric radii in the host galaxy, post-merger. Therefore, with our spatial cut, we tend to select these higher-metallicity stars from the progenitor.
On the flip side, for some of the halos where the GES sinks deeper into the center i.e., $\lesssim$3 kpc, our galactocentric radius cut removes the most-metal rich part of the GES population. Interestingly, previous works with the Auriga simulations (e.g., \citealt{orkney22}; \citealt{ciuca22}; \citealt{pinna23}), 
found that a starburst is induced during the MW-GES merger, increasing the overall stellar metallicity\footnote{By definition, these stars would be considered in-situ in this work.}. We show the relation between the average metallicity in the observable window, \fehsimwindow, vs the total population, \fehsimall, in Figure \ref{fig:mdf_1to1_line}. Although the \feh~of the GES population is different from the observable window compared to that from the total population, it is reassuring that there is still a positive trend--- i.e, the halos with higher \feh~for the total GES population generally have higher \feh~from the observable window as well, with an offset of $\sim 0.1-0.2$ dex. This difference therefore affects the estimated \mstar~from the [Fe/H] of GES stars in total versus those in the observable window, as later discussed in Section \ref{sec:sims_mstar_est}. In general, the \feh~in the observable window tends to be higher compared to the \feh~ of the overall population.
There are two halos where this is not the case, due to these progenitors being more massive and their higher-metallicity stars sinking deeper into the center of the Milky Way-like galaxy, which we exclude in our selection. 

\subsubsection{MDF from different selections}
\label{sec:mdf_sims_selections}

We now investigate the MDFs of GES from the observable window. Before this, however, we reiterate that the cuts we applied generally made the GES MDF closer to that of the in-situ population. And in fact, with the different selection methods explored, the MDFs have essentially been indistinguishable from each other. Therefore, separating the GES from the in-situ material based on their total MDF from each selection is nonsensical. Modeling this MDF to fit multiple components would produce artificial distributions that do not have physically motivated different stellar populations. 

This effect is largely due to the high-mass satellites in Auriga being too metal-rich (by $\sim$0.5 dex) compared to the observations, as has been 
shown in Figure 13 from \citet{grand21}. To alleviate this, 
we shifted the [Fe/H] of all tagged accreted material down by 0.5 dex, which we now call \fehsimshifted. In \citet{grand21}, the mismatch in the satellite MZR between the observations and simulations is stronger for GES-mass systems, which lessens and shows better agreement at the lower-mass end. However, we apply the 0.5 dex shift in [Fe/H] to all accreted population because the lower-mass systems contribute very minimally within the observable window, and even more so in the selection methods explored. We use \fehsimshifted~in identifying the MDF of the GES vs in-situ stellar populations. 

With the \fehsimshifted, we show the normalized (un-normalized) MDFs for the different GES selections in Au-18 as the grey solid histograms in the top row (bottom row) of Figure \ref{fig:mdf_3selections_sims}. We perform GMM on these MDFs to determine the number of different components that contribute to the distribution.
The optimal number of components was determined through BIC, and these Gaussians are shown as the dashed lines in the first row of Figure \ref{fig:mdf_3selections_sims} (the total best-fit GMM is shown with the solid line). In determining the peak that is associated with the GES, we are informed by our exploration in Section 
\ref{sec:kinematic_sims}, specifically pertaining to the purity of the selection summarized in Table \ref{tab:pure_comp}. This tells us that in the simulations, the majority of the selected GES stars from any of the methods \textit{are not} associated with GES. However, of all the accreted material, the GES merger is typically the most dominant contributor in the halo at these distances (see also \citealt{fattahi19} regarding how this sample of Auriga halos was selected). 

\begin{figure}
    \centering
    \includegraphics[width=0.9\columnwidth]{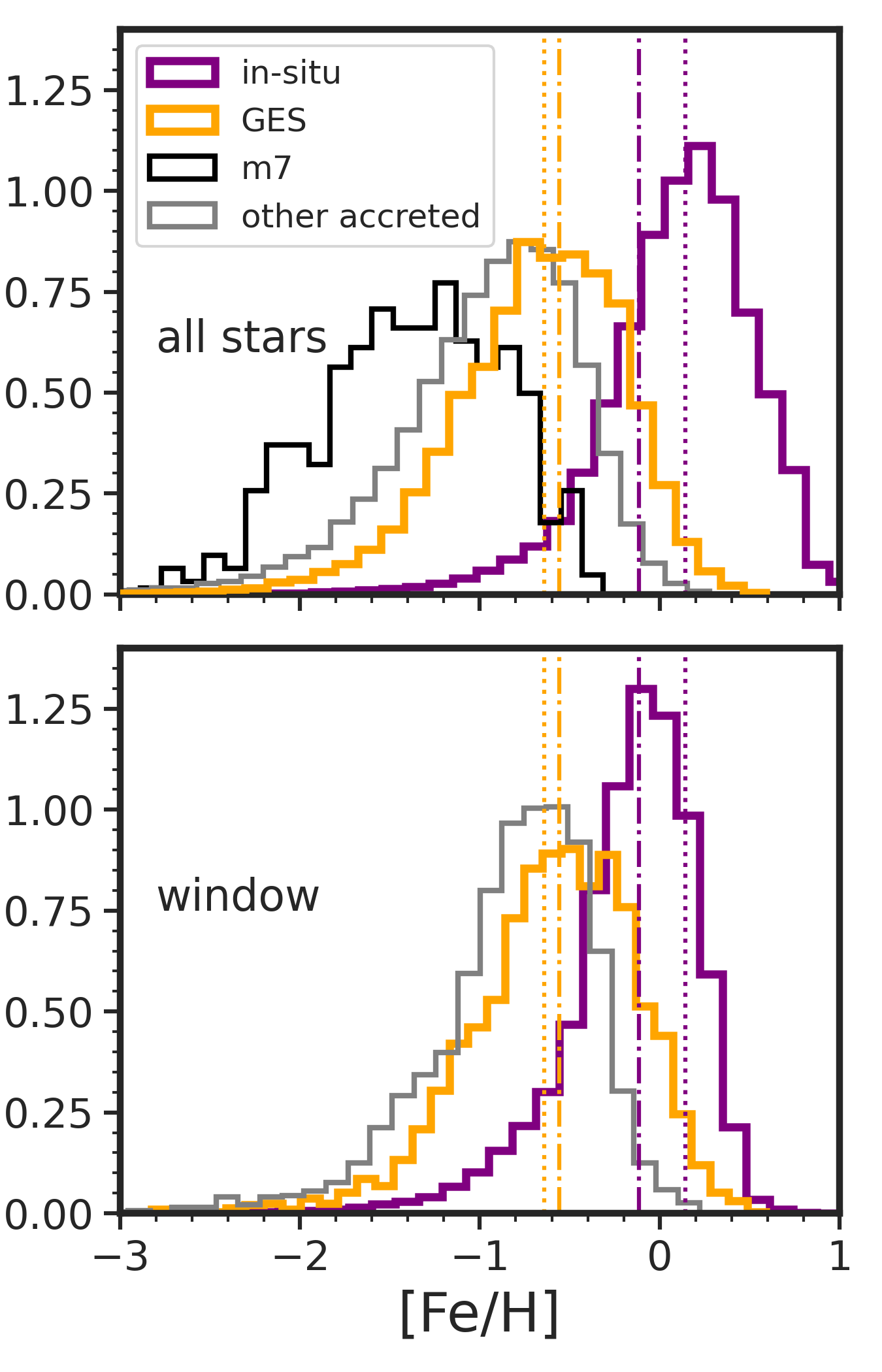}
    \caption{\textit{Metallicity distribution functions for the total population and in the observational window.} \textit{Top:} MDF of the total populations for GES (orange), in-situ stars (purple), a lower mass system with \mstar~$= 1.4 \times 10^{7} \rm M_{\odot}$ (m7, black), and the rest of the accreted material (grey) for Au-18. The MDFs peak at higher [Fe/H] for higher mass systems. \textit{Bottom:} MDF for the GES (orange), in-situ (purple), and the rest of the accreted (grey) stars in our defined observational window for Au-18. Both panels include the median [Fe/H] from the total population (dotted line) and in the observational region (dot-dash line) for the in-situ stars and GES. The MDFs of the in-situ stars and GES seem much closer to each other due to the removal of the higher-metallicity disc stars and some lower-metallicity GES stars. The MDFs are normalized such that the area under the curve equals one.}  
    \label{fig:mdf_sims_gse_ins}
\end{figure}

\begin{figure}
    \centering
    \includegraphics[width=0.9\columnwidth]{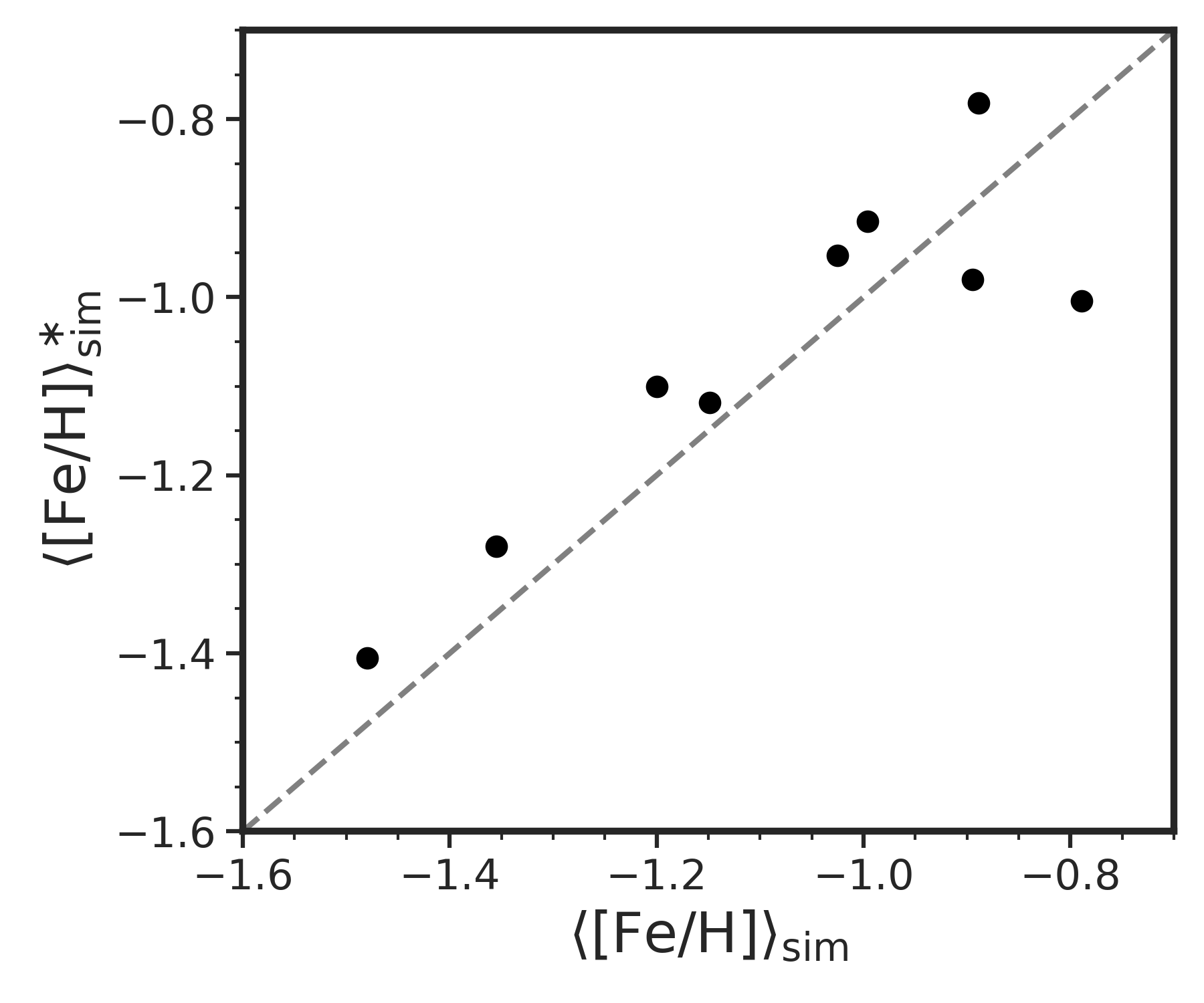}
    \caption{Comparison of the metallicity for the total GES population (x-axis, \fehsimall) vs the GES stars in the observable window (y-axis, \fehsimwindow). Dashed grey line shows 1:1 correspondence. The majority of the GES material in the observable window has higher [Fe/H] than the total population. Due to the progenitor's negative metallicity gradient, i.e., its central region tends to have higher metallicity stars than the outskirts, and our spatial selection of $\lesssim$20 kpc, we miss the lower metallicity tail of the progenitor that are at greater distances.}
    \label{fig:mdf_1to1_line}
\end{figure}

\begin{figure*}
    \centering
    \includegraphics[width=0.9\textwidth]{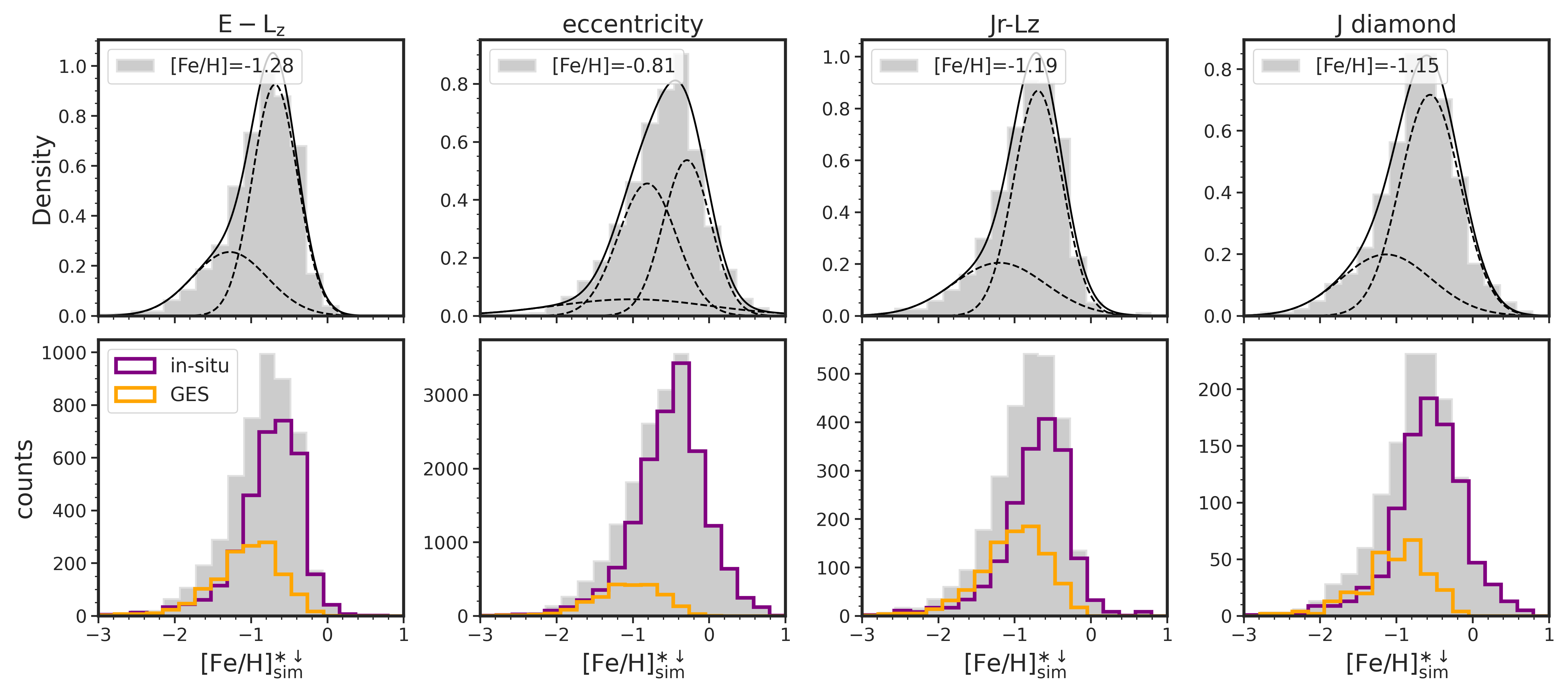}
    \caption{\textit{Metallicity distribution functions from different selection methods in the simulations.} \textit{Top:} Normalized MDF of the stellar populations from the different selection methods in the simulations. From left to right, we show the samples from the E-Lz, eccentricity, Jr-Lz, and action diamond selections. Black line indicates the total best-fit from the GMM while the solid dashed lines are the individual Gaussians. The associated GES [Fe/H] determined from the GMM is noted in the legend of each subpanel. \textit{Bottom: } Un-normalized MDF of the stellar populations from the selection methods showing the true in-situ (purple) and GES (orange) populations within each sample. The metallicity shown here is \fehsimshifted~i.e., the [Fe/H] in the observable window, with the accreted material shifted down by -0.5 dex. See Sections \ref{sec:mdf_sim_allvobs} and \ref{sec:mdf_sims_selections} for this discussion. 
    \label{fig:mdf_3selections_sims}}
\end{figure*}

With this in mind, we ascertained that it is the second most dominant component (which turns out to also be the second highest-metallicity component) that is associated with GES. For example and as shown in Figure \ref{fig:mdf_3selections_sims}, the GMM for the MDF of Au-18 gives a GES that peaks at \fehsimshifted$=-$1.28, -0.81, -1.19, and -1.15 for the E-Lz, eccentricity, Jr-Lz, and action diamond selections, respectively. On the other hand, the actual, most dominant component for each selection method in the simulations corresponds to the in-situ material which peaks at a higher [Fe/H]. We tabulate the different mean \fehsimshifted~for GES from each selection in Table \ref{tab:auriga_mdf} for all nine halos investigated. We note again that we are quoting \fehsimshifted~in the table, which are shifted lower by 0.5 dex from the true value in the observational window in the simulations, \fehsimwindow. Comparing the MDFs in the observations (Figure \ref{fig:MDFs_4selection}) to these ones in the simulations (Figure \ref{fig:mdf_3selections_sims}) highlights that the latter also look different in that they are not multi-modal. The in-situ stars in the Milky Way have a more widely separable MDF from the GES stars in the observations, whereas they are less distinguishable in the simulations, as illustrated in Figure \ref{fig:mdf_sims_gse_ins}. We note that these two effects i.e., the in-situ material dominating the selection instead of GES stars, and the GES and in-situ MDFs being less distinguishable from each other, are due to the following factors: (1) the larger stellar masses for the GES-like systems in Auriga especially with respect to the Milky Way host enable the galaxies to be more chemically enriched than is expected in the observations, (2) at a similar mass scale, the satellite galaxies in Auriga generally have higher [Fe/H] than seen in the observations, (3) the in-situ discs in Auriga are different and thicker than the real Milky Way's, therefore the spatial cut we applied is not getting rid of a lot of in-situ stars as we do in the observations, and (4) none of these are ``exact" GES in terms of their mass and orbit so the ratio of in-situ to GES stars can vary and will not be exactly like what we see in the Milky Way. 
Therefore, we expect the contamination in the simulations to be different and in fact much larger compared to the true contamination in the observations. Nonetheless, exploring these halos in the simulations is informative as they still correspond to a massive merger that happened at earlier epochs, and contribute significantly to the radially anisotropic stellar population in the halo, much like the GES in the observations. 

We also show the MDFs for the true in-situ and GES populations from each selection in the bottom row of Figure \ref{fig:mdf_3selections_sims}. This confirms our assignment of which Gaussian component corresponds to GES, aside from the Jr-Lz selection which was best-fit to have three components, because of (1) the lower [Fe/H] peak of GES compared to the in-situ material and (2) the larger spread in the MDF of GES stars.  
One obvious drawback in using GMM to separate the in-situ vs GES MDFs is also highlighted here: the true in-situ and GES MDFs do not necessarily follow a Gaussian distribution, though we have applied this assumption. Nonetheless, it is reassuring that the GES (orange histogram) and in-situ (purple histogram) populations make up the bulk of the sample in each selection method, as adding those two histograms together would give the total MDF (gray histogram). 

\subsection{Stellar mass determination}
\label{sec:sims_mstar_est}

\begin{figure}
    \centering  \includegraphics[width=0.98\columnwidth]{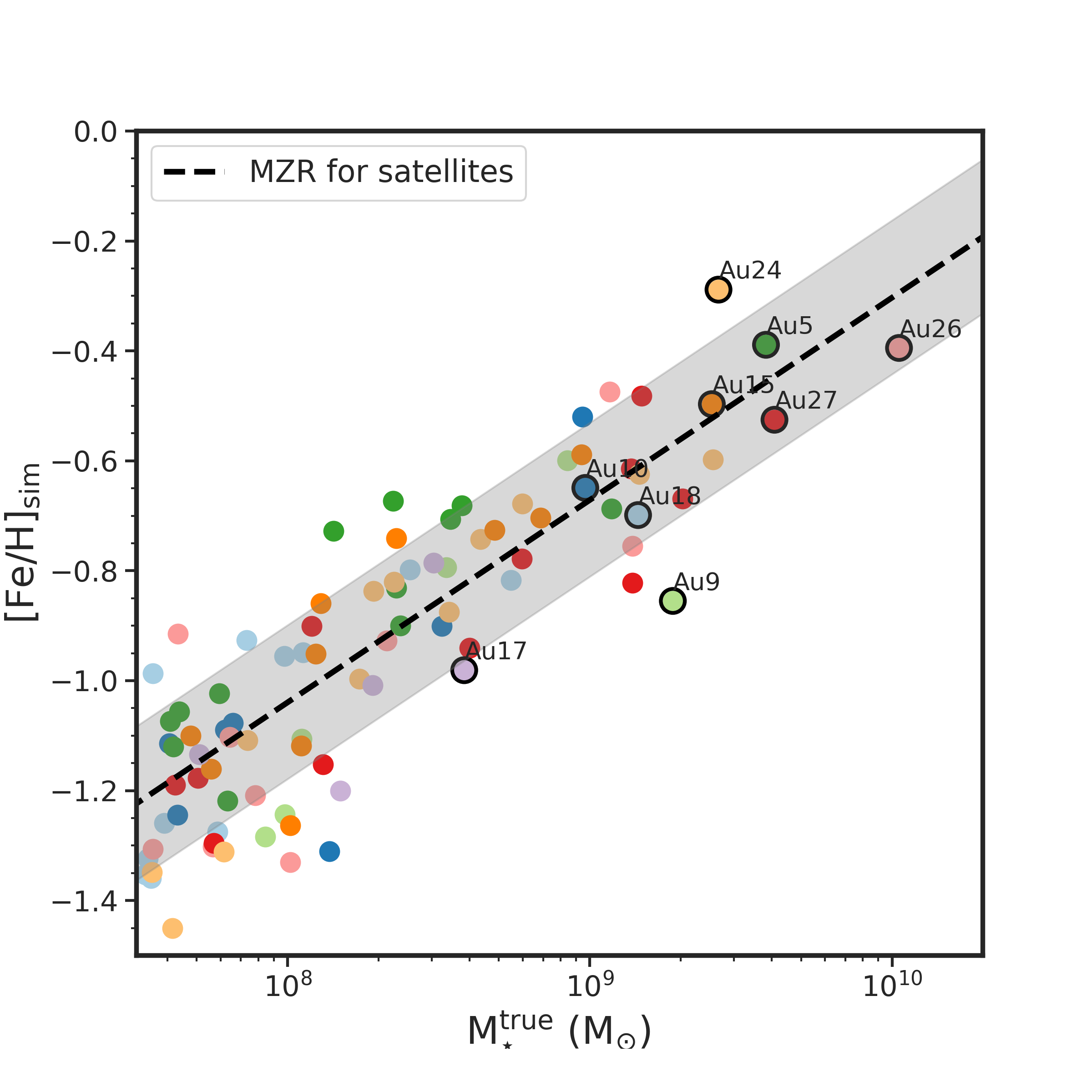}
    \caption{Mass-Metallicity relationship (MZR) in the satellites around the Auriga halos that have GES-like mergers. Every color corresponds to a different satellite population around the same central galaxy. The GES in each halo is highlighted with a black circle. The MZR in the simulations is shown as a dashed black line with one sigma above and below this relation filled in grey.}
    \label{fig:mzr_sims}
\end{figure}

In the previous section, we have thoroughly explored the MDF of the GES population---as a whole (e.g., \fehsimall), in the observable window (e.g., \fehsimwindow~and \fehsimshifted which is shifted down by 0.5 dex), and with the different selection methods. Now we use the [Fe/H] estimates from the different GES selections to convert to a stellar mass estimate as we have done in the observations.  In Figure \ref{fig:mdf_sims_gse_ins}, there appears to be a mass-metallicity relationship in the satellites around the Milky Way systems in Auriga with the MDFs arranged from lowest to highest metallicity going from the least to the most massive system. 

We further explore this relationship and show the mean \fehsimall~vs peak stellar mass, \mstar, for the destroyed satellites around the Auriga halos with GES-like mergers in Figure \ref{fig:mzr_sims}. 
It is apparent that the MZR in the simulations holds for a large range in \mstar~and [Fe/H], specifically over the range where the GES progenitors lie. This has been similarly noted in previous studies on the Auriga satellites (e.g., \citealt{grand21}, \citealt{deason23}). We adopted a simple linear relationship to describe the MZR, given by \feh~~$=$~0.37~$\times$~log(\mstar/$\rm M_{\odot}$) - 3.99 with a scatter of 0.13~dex/log(\mstar/$\rm M_{\odot}$). 

\begin{figure}
    \centering  \includegraphics[width=0.98\columnwidth]{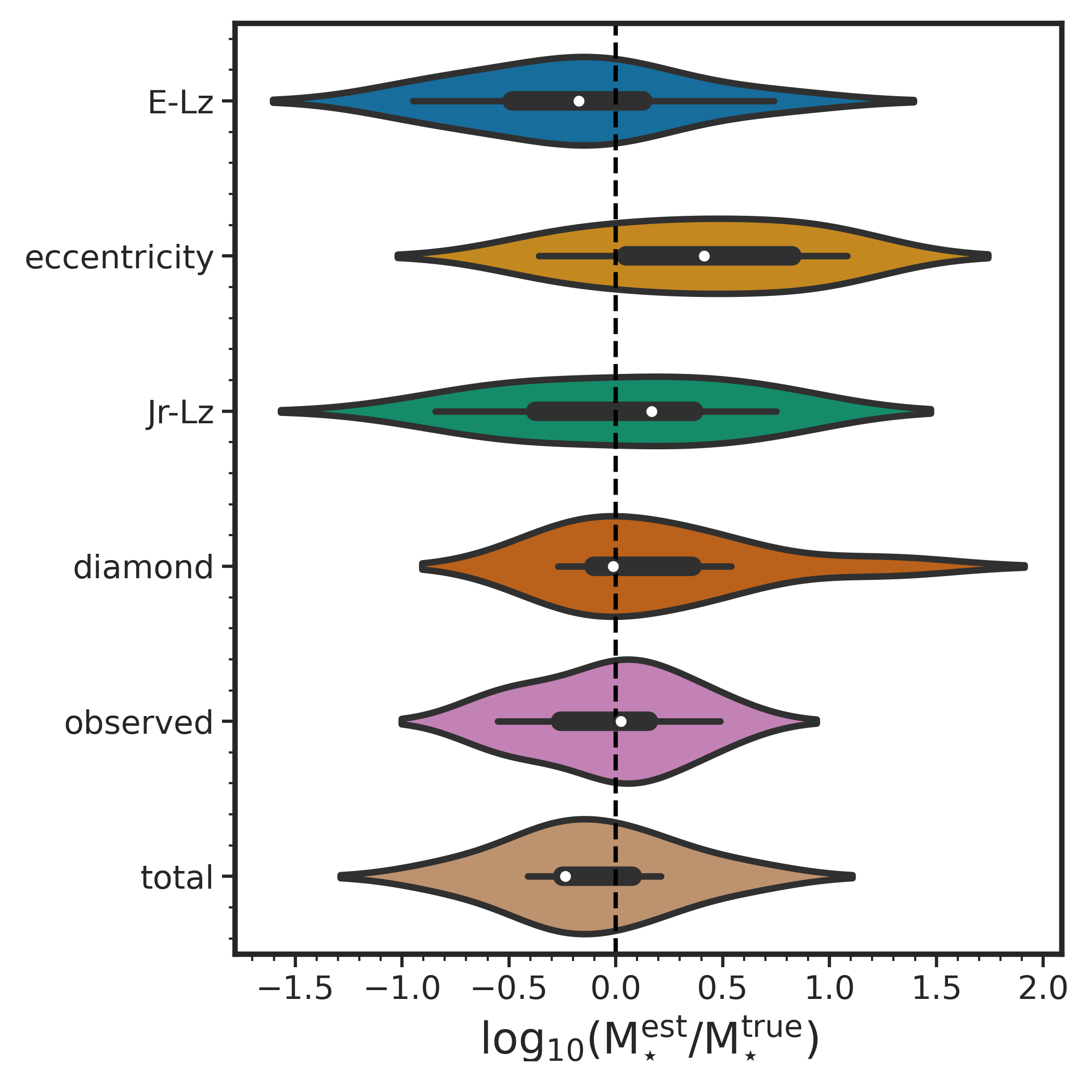}
    \caption{Violin plot of mass estimate ratios derived from [Fe/H] for the E-Lz, eccentricity, Jr-Lz, and action diamond methods as well as GES population in the observable window and the GES total population, ordered from top to bottom. }
\label{fig:mratio_violin}
\end{figure}

We adopt this MZR to get  the stellar mass estimates, \mstarest, for the GES-like systems in Auriga listed in Table \ref{tab:auriga_mdf}.  We shifted back up the \fehsimshifted~ in Table \ref{tab:auriga_mdf} by 0.5 dex to \fehsimwindow~and used this metallicity to derive \mstarest~ such that the GES metallicities are back on the same scale as the MZR in the simulations. 
We illustrate the distribution of the different mass estimates for the nine Auriga halos in Figure \ref{fig:mratio_violin}. Specifically, we show a violin plot of the $\rm log_{10}$(\mstarest/\mstartrue) from the different selection spaces to more easily see their distribution with respect to the \mstartrue~(marked as dashed line). The white dot shows the median, the grey bar shows the first and third quartiles, and the thin grey line shows the rest of the distribution, barring outliers. In general, all mass estimates are able to reproduce the \mstartrue~but to varying precision and accuracy. We note that again, a lot of the scatter here is mostly due to the different assembly histories of the halos that we investigated. However, this is also why comparisons to the real GES stars in the observational window (pink) and its total population (brown) are elucidating. 

Interestingly, the \mstarest~derived from the GES stars in the observational window seems to be closer to the true value compared to if we take \textit{all} of the GES stars. The former has the median value right at the true value with \logmassratio$=$0.02 while the latter is underestimated with \logmassratio$=-0.24$. However, based on Figure \ref{fig:mzr_sims}, more than half (5/9) of the GES in the simulations lie below the MZR. The offset between the MZR and the progenitors below this relationship is also larger compared to the offset with the progenitors that lie above it. The underestimated mass from the total population is therefore expected, while the ``correction" to the true value from the observational window is due to the bias towards the higher metallicity stars of the GES progenitor. We note however that this distribution of the \mstarest~from the total population happened by chance, purely based on where they lie on the MZR.   

In terms of accuracy on \mstartrue, the action diamond (orange) method is the most accurate with median \logmassratio$=-0.01$. The \mstarest~distributions from the E-Lz (blue) and Jr-Lz (green) methods are also generally accurate, i.e., crossing the 1-to-1 line, though in bulk the E-Lz method underestimates the true mass with \logmassratio$=-0.17$ (factor of 0.7 lower) and Jr-Lz overestimating with \logmassratio$=0.17$ (factor of 1.5 higher). The eccentricity method is the least accurate with a median \logmassratio$=0.41$, which overestimates the \mstartrue~by a factor of 2.6.
Although the eccentricity selection is the most complete, it also has the lowest purity. The fact that it is the least accurate in estimating the true \mstar~is in line with this. The converse is not true however; the purest selection, Jr-Lz, does not necessarily give the most accurate \mstarest~but the action diamond does. This is also interesting as the action diamond by far has the lowest completeness of all dynamical selections and it has a comparable purity to the E-Lz selection. We do want to emphasize that these distributions are based only on nine GES-like systems and are therefore prone to small number statistics. Nonetheless, it seems that a smaller GES sample size and a purer selection given by the Jr-Lz and action diamond methods, lend to more accurate stellar mass estimates. Meanwhile, a larger GES sample that is more complete but less pure such as that from the eccentricity method is the least accurate and overestimates the stellar mass.  


This exploration of nine GES-like progenitors in the Auriga simulations has shown that each selection method is beneficial in their own right. The selection in Jr-Lz wins in terms of purity (33\%), eccentricity in completeness (62\%), and the E-Lz, Jr-Lz, but most of all the action diamond, in inferring \mstartrue. Interestingly, though unsurprisingly, the \feh~for the total GES population is lower compared to the \feh~for the GES population in the observational window because we are probing populations that were more centrally located in the progenitor pre-merger, which are also higher in [Fe/H]. 
Although the \mstar~estimate from the MZR for the total population is generally underestimated because the majority of the halos happened to be below the MZR, the bias towards higher [Fe/H] in the observational window brings the \mstar~estimate closer to the true value.  


\begin{table*}
\begin{center}
\caption{Mean [Fe/H] of the GES component from the GMM in the observable window, and the associated \mstar~estimate for GES for the different selection methods in the simulations. The [Fe/H] listed in columns 2-5 are \fehsimshifted, i.e., shifted down by 0.5 dex which was applied during the GMM (see Section \ref{sec:mdf_sims_selections}). The stellar masses in columns 6-9 are derived from shifting back up the listed metallicities by 0.5 dex and using the MZR for the satellites in the simulations.}
\begin{tabular}{@{\extracolsep{8pt}}ccccccccc}
\hline \hline
(1) & (2) & (3) & (4) & (5) & (6) & (7) & (8) & (9) \\
halo & \fehsimshifted & \fehsimshifted & \fehsimshifted & \fehsimshifted & $\rm log_{10} (M_{\star}^{est})$ & $\rm log_{10} (M_{\star}^{est})$ & $\rm log_{10} (M_{\star}^{est})$ & $\rm log_{10} (M_{\star}^{est})$ \\
& E-Lz & eccentricity & Jr-Lz & J diamond & E-Lz & eccentricity & Jr-Lz & J diamond \\
\cline{2-5} \cline{6-9}
\\
Au-5 & -1.00 $\pm$0.11 & -1.09 $\pm$0.08 & -0.89 $\pm$0.05 & -0.99 $\pm$0.04 & 9.45 $\pm$0.50 & 9.27 $\pm$0.43 & 9.78 $\pm$0.42 & 9.48 $\pm$0.40 \\
Au-9 & -1.01 $\pm$0.24 & -1.04 $\pm$0.39 & -0.93 $\pm$0.07 & -1.17 $\pm$0.28 & 9.43 $\pm$0.76 & 9.33 $\pm$1.13 & 9.64 $\pm$0.41 & 9.07 $\pm$0.87 \\
Au-10 & -0.90 $\pm$0.24 & -1.09 $\pm$0.35 & -1.10 $\pm$0.26 & -0.97 $\pm$0.12 & 9.71 $\pm$0.76 & 9.24 $\pm$0.99 & 9.18 $\pm$0.84 & 9.52 $\pm$0.49 \\
Au-15 & -0.96 $\pm$0.34 & -0.63 $\pm$0.08 & -0.75 $\pm$0.07 & -0.89 $\pm$0.20 & 9.55 $\pm$0.99 & 10.45 $\pm$0.43 & 10.15 $\pm$0.44 & 9.79 $\pm$0.67 \\
Au-17 & -1.42 $\pm$0.31 & -1.19 $\pm$0.31 & -1.04 $\pm$0.13 & -0.84 $\pm$0.16 & 8.26 $\pm$0.93 & 8.97 $\pm$0.90 & 9.39 $\pm$0.54 & 9.92 $\pm$0.55 \\
Au-18 & -1.28 $\pm$0.24 & -0.81 $\pm$0.14 & -1.19 $\pm$0.36 & -1.15 $\pm$0.35 & 8.71 $\pm$0.75 & 9.98 $\pm$0.54 & 8.93 $\pm$1.09 & 9.05 $\pm$1.03 \\
Au-24 & -1.09 $\pm$0.19 & -0.81 $\pm$0.07 & -1.22 $\pm$0.20 & -0.90 $\pm$0.09 & 9.21 $\pm$0.64 & 9.98 $\pm$0.44 & 8.87 $\pm$0.66 & 9.73 $\pm$0.46 \\
Au-26 & -1.15 $\pm$0.07 & -0.87 $\pm$0.07 & -1.11 $\pm$0.18 & -0.90 $\pm$0.10 & 9.01 $\pm$0.41 & 9.81 $\pm$0.41 & 9.18 $\pm$0.63 & 9.72 $\pm$0.45 \\
Au-27 & -1.22 $\pm$0.25 & -0.59 $\pm$0.03 & -1.09 $\pm$0.14 & -0.96 $\pm$0.11 & 8.86 $\pm$0.79 & 10.56 $\pm$0.40 & 9.24 $\pm$0.53 & 9.63 $\pm$0.49 \\
\hline
\hline
\end{tabular}
\label{tab:auriga_mdf}
\end{center}
\end{table*}

\section{Discussion}
\label{sec:discussion}



In the first part of this work focusing on the observations (Section 
\ref{sec:observation_data}), we have determined that the estimated stellar mass of GES differs by a factor of $\sim$2 depending on the selection method. The [Mg/Mn]-[Al/Fe] selection gives the lowest estimate while the eccentricity selection gives the highest estimate based on their MDFs. The assumption that largely changes the \mstar~however is the adopted MZR which ranges from $2.37 - 5.26 \times 10^{8}$ $\rm M_{\odot}$ assuming the $z=0$ \citet{kirby13} relation to $2.45 - 5.30 \times 10^{9}$ $\rm M_{\odot}$ assuming the disrupted dwarfs ($z>$0) relation from \citet{naidu22}. This is an order of magnitude difference! Suffice to say, the real \mstar~of GES is likely in between these values. 

In the second part focusing on the simulations, we then test our method of identifying GES stars and similarly derive their \mstar~using the MZR for the disrupted satellites in Auriga. We are then able to compare this to the true value in the simulations shown in Figure \ref{fig:mratio_violin}. 
There is not one method that gives the highest (or lowest) estimate for the \mstar~across \textit{all} the Auriga halos listed in Table \ref{tab:auriga_mdf} because of their different assembly histories.
However, in general, the action diamond method gives the most accurate \mstarest, while the eccentricity method is the least accurate and on average overestimates the \mstar~by a factor of $\sim$2.6. In the observations, the eccentricity method similarly gives the highest \mstarest~for GES, followed by the action diamond, E-Lz, Jr-Lz, and lastly [Mg/Mn]-[Al/Fe]. The GES system in Au-24 follows a similar trend in the \mstarest~as in the observations, which has a peak \mstartrue$= 2.56~\times~10^{9} \rm M_{\odot}$.  From Table \ref{tab:auriga_mdf}, the variation in \mstarest~\textit{within} each Auriga halo is larger in the simulations compared to that from the observations. We conjecture that this is driven by the GES being more massive and having higher metallicity in the simulations, making it less distinguishable from the in-situ population compared to the observed data and therefore affecting the [Fe/H] from which we derive a stellar mass.


For the majority of this work, we have gone through the details of deriving the \mstar~of GES. But for understanding whether or not we can pick and choose a GES selection method, it is worth looking into other progenitor properties as well.


\subsection{GES eccentricity}
\label{sec:gse_ec_met}

\begin{figure*}
    \center  
    \includegraphics[width=0.88\textwidth]{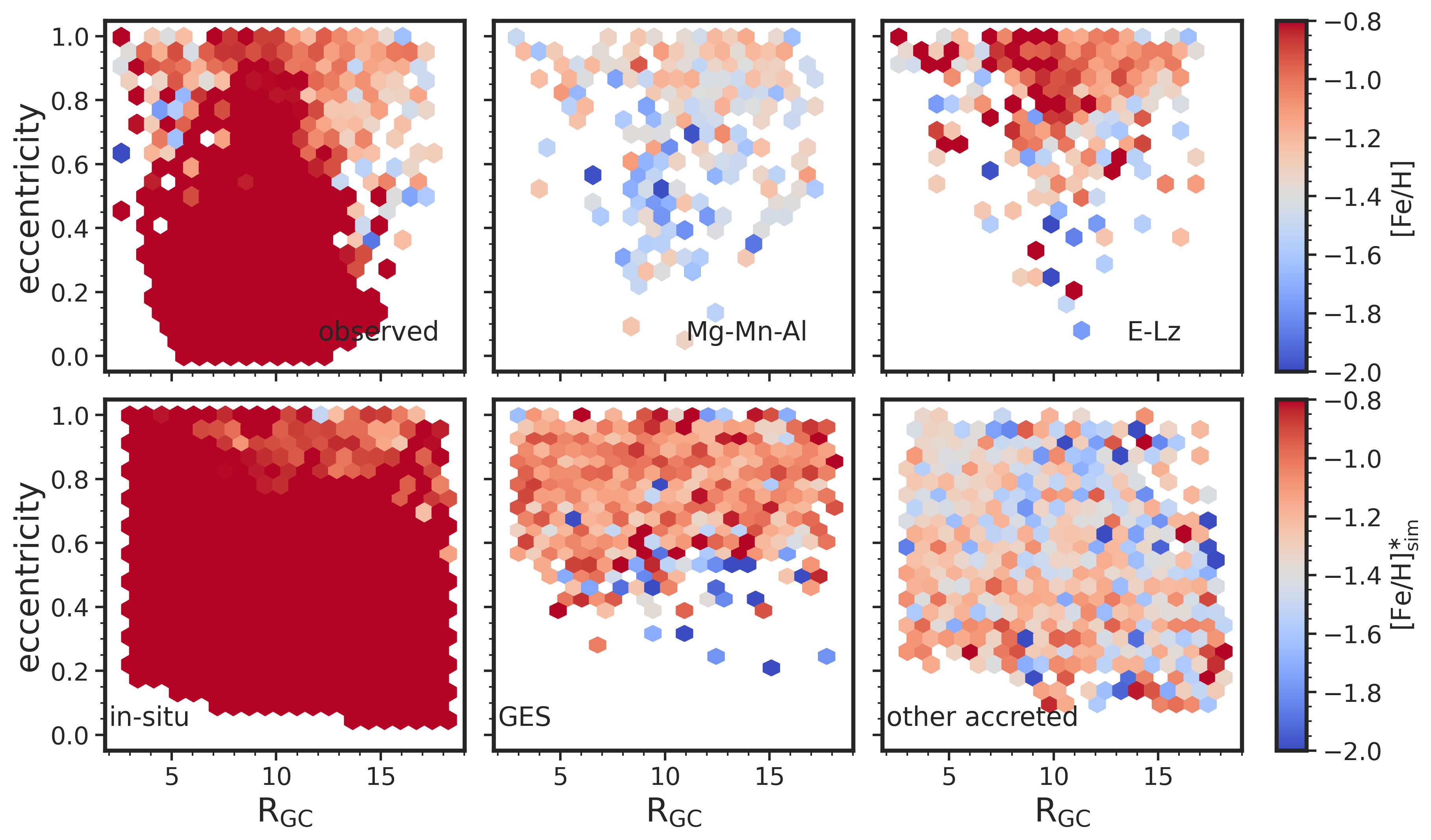}
    \caption{\textit{Eccentricity vs current Galactocentric radius, $\rm R_{GC}$, colored by metallicity.} The top row shows this for all the observed data as well as the GES samples through the [Mg/Mn]-[Al/Fe] and E-Lz selection methods (from left to right, respectively). The bottom row shows the in-situ, GES, and all other accreted material in the observational window for Auriga 18. It is noteworthy that the in-situ stars have decreasing metallicity with higher eccentricity while the GES shows the opposite trend.
    The two selection methods in the observations qualitatively match our expectations from the simulations. }
\label{fig:ecc_v_R}
\end{figure*}

We further check the validity of our selections by looking at the eccentricity of stars versus their current Galactocentric radius, $\rm R_{GC}$, as shown in Figure \ref{fig:ecc_v_R}. The top row shows these properties as seen in the observations for all the stars (left), and the GES stars as selected from [Mg/Mn] vs [Al/Fe] (center), and from E-Lz (right). We only show results from these two selection methods because they are the only ones that span a large range in eccentricity (see Figure \ref{fig:ges_chem_4panels}).
The bottom row shows the same plot from the observable window for stars that are in-situ (left), from GES (center), and from all other accreted material (right) for Au-18 in the simulations. In addition, both the top and bottom rows are colored by [Fe/H] but we note that the ranges are different between the observations and the simulations.

It is apparent that the way in-situ stars occupy this space is different compared to the GES stars. Firstly, the parent sample in the observations is clearly dominated by higher-metallicity in-situ stars that in general show progressively lower metallicities with higher $\rm R_{GC}$.
This is similarly observed for the in-situ stars in the simulations as well. One difference however, is that the lower $\rm R_{GC}$ sample in the observations show lower metallicities at high eccentricity whereas this does not exist in the simulations. This is likely due to the differences in the Auriga discs vs the Milky Way in addition to our simple selection function in the simulations. In contrast, the true GES stars in the simulations do not show such a progression which at first glance seems to counter our intuition about the presence of a metallicity gradient. However, this is because we are mainly looking at their current $\rm R_{GC}$ (over a small range in $\rm R_{GC}$) which is not a preserved quantity. Of the two selection methods investigated here, the chemistry selection resembles GES better as suggested by the simulations. The E-Lz selection shows a metallicity gradient with higher [Fe/H] at lower $\rm R_{GC}$ which is due to contamination from in-situ stars as shown in Figures \ref{fig:ges_chem_4panels} and \ref{fig:MDFs_4selection}. 

Another interesting distinction between the in-situ and GES stars is the opposite [Fe/H] gradient from low to high eccentricity. For the in-situ stars, the lower eccentricity stars have higher [Fe/H] which progressively goes to lower [Fe/H] at higher eccentricity, especially at higher $\rm R_{GC}$. Indeed we see this both in the observations and the simulations. On the other hand, it is the opposite story for the GES stars---at lower eccentricity, though there are fewer stars in this region, the GES stars have lower [Fe/H] \footnote{We have investigated this diagram for the other Auriga halos in this work and they in fact show stronger trends. However, we choose to show Au18 for consistency.}. A larger percentage of the higher [Fe/H] sample are in fact at higher eccentricities as well. This is reasonable if we consider a pre-existing negative metallicity gradient for the GES progenitor and that the stars that are stripped later are more centrally located and highly radialized due to dynamical friction \citep{amorisco17,vasiliev19,amarante22}. Indeed, we similarly see this in the observations for both selection methods. However, due to the contamination from in-situ stars in the E-Lz method, this [Fe/H] gradient along the eccentricity axis seems to be stronger compared to that from the [Mg/Mn] vs [Al/Fe] method and from the true GES population in Au-18. 

We also show this eccentricity vs $\rm R_{GC}$ diagram for all other accreted material in the observable window in Au-18 to highlight that this looks different from the GES stars. For all other accreted material, the [Fe/H] gradient is opposite to the trend for GES as a function of eccentricity. That is, all other accreted stars have decreasing [Fe/H] with higher eccentricity. In addition, at the high eccentricity region, the [Fe/H] for the other accreted material are lower than the GES. 

With this exploration, we find that the chemical and E-Lz selections qualitatively reproduce the trends in eccentricity, $\rm R_{GC}$, and [Fe/H] as the true GES population. However, due to the larger contamination of in-situ stars in the E-Lz selection, the chemical selection is the better method for encompassing the nature of the GES along these axes. 


\subsection{The total mass of GES}
\label{sec:gse_totmass}

\begin{figure}
    \centering  
    \includegraphics[width=\columnwidth]{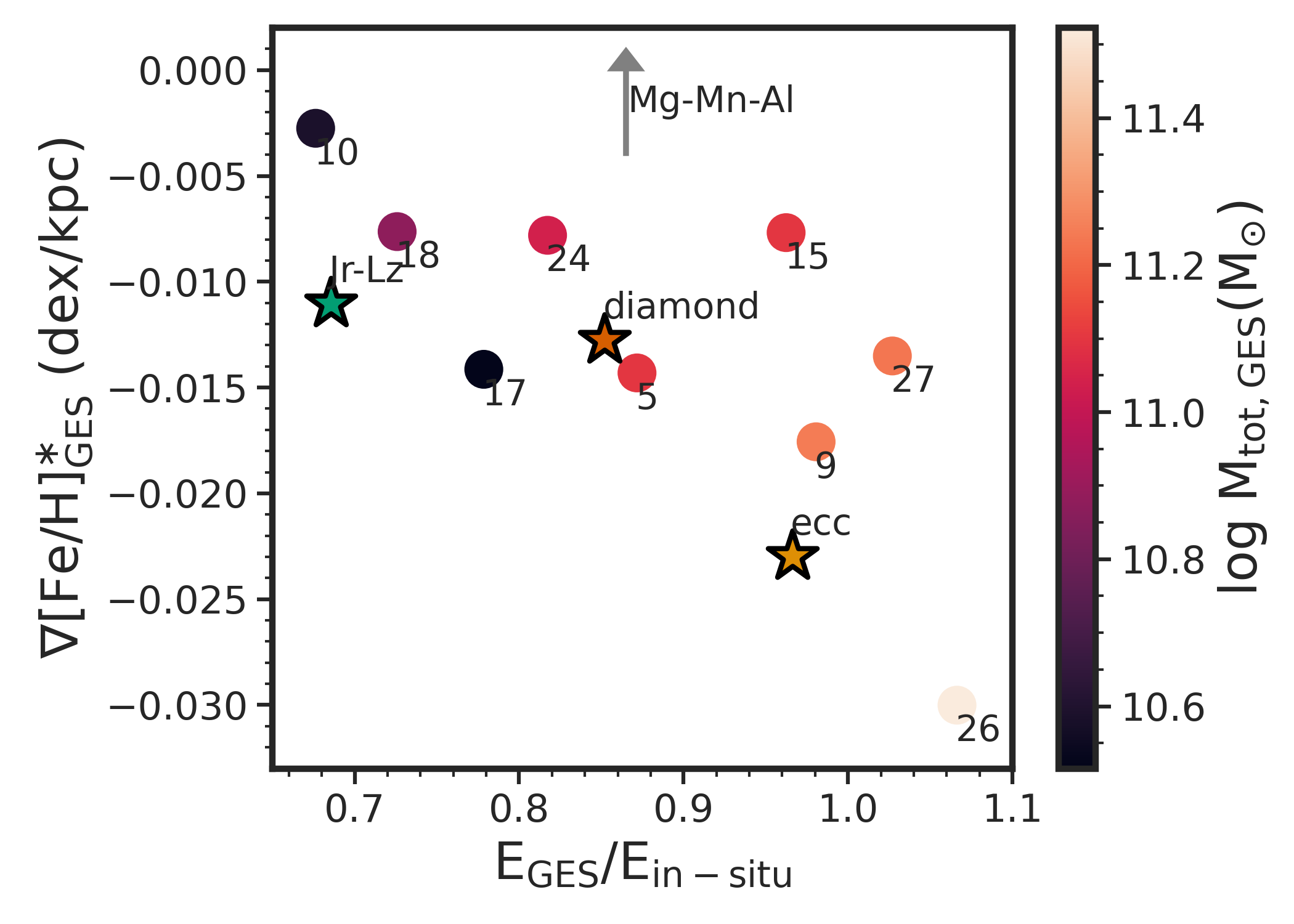}
    \caption{\textit{Metallicity gradient versus energy of Auriga GES systems in the observational window.} The metallicity gradient, \fehgrad, is obtained using the mean orbital radius i.e., the average between the apocentre and pericentre of the star. $\rm E_{GES}/E_{in-situ}$ is defined as the ratio between the 10th percentile of the GES stars and the 10th percentile of the in-situ stars in energy space. The points are also colored by the GES total mass. In addition, we show the GES from the Jr-Lz (green), action diamond (orange), and eccentricity (yellow) selections in the observations as stars. The data point for the [Mg/Mn]-[Al/Fe] selection is outside this diagram and marked with an arrow. }
\label{fig:fehgrad_esink_mass}
\end{figure}

So far, we have been looking at the stellar mass of GES---now we will focus on obtaining the total mass using only observational quantities. In Figure \ref{fig:fehgrad_esink_mass}, we show the metallicity gradient for the GES stars versus how deep the GES sinks into the potential for the nine halos we looked at in Auriga. Here, we use only the stars that are within the observational window in the simulations. In addition, we colored these circles by the total mass of GES  in Auriga. The metallicity gradient, \fehgrad, is obtained using the mean orbital radius i.e., the average between the apocentre and pericentre of the star, which should retain more pre-merger information compared to the current $\rm R_{GC}$. To quantify how deep the GES sinks into the potential, we use the in-situ stars as a reference point. We took the ratio of the 10th percentile of the GES stars and the 10th percentile of the in-situ stars in energy space, $\rm E_{GES}/E_{in-situ}$. The relative trend is more important than the absolute values here---GES systems with lower ratios (to the left) lie at higher energies while those with larger ratios are at lower energies and are more bound. 

By comparing these two observables in the simulations we already see a clear trend: systems with a steeper metallicity gradient (i.e., more negative) sink deeper into the potential (i.e., higher $\rm E_{GES}/E_{in-situ}$). This is made even clearer when looking at the total mass of GES. Indeed, the systems that sink deeper and have steeper metallicity gradients correspond to halos that are more massive. This is expected as massive systems would have stars that are more bound to the progenitor, which would therefore be stripped much later when GES is deeper into the host's potential. It also makes sense that these same systems have steeper gradients as \citet{monachesi19} found that the Auriga stellar halos that have steeper metallicity gradients tend to be dominated by a few significantly large mergers.

With this promising result, we can potentially use the metallicity gradient and $\rm E_{GES}/E_{in-situ}$ of our selected GES stars in the observations to obtain a total mass of the GES progenitor. We test this out for the GES stars selected through the chemistry, eccentricity, Jr-Lz, and action diamond methods shown as the non-circular markers in Figure \ref{fig:fehgrad_esink_mass}. Because our selection function in the simulations is very simple, we are not mimicking the fact that we should observe fewer stars with lower $\rm R_{GC}$ (that are at larger heliocentric distances) because they are fainter. These stars are more bound and have lower energies which increase the absolute value in the denominator of $\rm E_{GES}/E_{in-situ}$. This is \textit{not} what we see for the parent sample in the observations (dominated by in-situ stars) because we are biased by the higher energies and larger Lz of the brighter disc stars. Therefore, we only look at stars with [Fe/H] $<$ -0.8 in calculating $\rm E_{GES}/E_{in-situ}$ to mitigate the effect of the higher energy stars from the disc. On the other hand, we note that we are potentially losing the most bound and highest metallicity stars from GES by imposing this. In addition, although we apply this cut in calculating how deep the GES sinks into the Milky Way potential, we do not apply the same cut for calculating the metallicity gradient as this would introduce a sharp and artificial boundary.

The Jr-Lz, action diamond, and eccentricity selected GES samples have similarly negative metallicity gradients as the GES in the simulations. They in fact lie on the same metallicity gradient-energy relation which is reassuring. In contrast, the chemically-selected GES sample has a positive metallicity gradient. Upon inspection, we find that this is largely due to a metallicity bias in that this sample has the lowest value in [Fe/H] in the innermost mean orbital radius i.e., [Fe/H]$\sim$-1.0 compared to the other selections that reach [Fe/H]$\sim$-0.5 at the same radius. This had the net effect of inverting and making the \fehgrad~positive for the chemical selection. We are more likely to believe the metallicity gradient obtained from the Jr-Lz method because (1) it is less contaminated by in-situ material in the observations as shown in Figures \ref{fig:ges_chem_4panels} and \ref{fig:MDFs_4selection}, and (2) this is backed by the simulations that show it selects the purest GES sample (see Table \ref{tab:pure_comp}). The action diamond selection also seems promising and gives \fehgrad~and $\rm E_{GES}/E_{in-situ}$ values in between the Jr-Lz and eccentricity methods.

From where the GES lies based on the Jr-Lz and action diamond methods, we estimate that \textit{GES has a total mass of $10^{10.5 - 11.1}~M_{\odot}$} in the same range as that of Au5, Au10, Au17, Au18, and Au24. This independent estimate is lower but well within the range of halo masses obtained using the SMHM relation from \citet{behroozi19} in Table \ref{tab:MDF}. This also agrees with previous estimates of the GES total mass using tailored N-body simulations (e.g., \citealt{naidu21b}) and globular clusters (e.g., \citealt{callingham22}). This method is indeed promising because the total mass estimate bypasses any assumptions on the form of the stellar mass-metallicity relationship and the SMHM relation. In addition, it does not rely on the absolute values of GES properties but instead deals with the \textit{relative trends} in [Fe/H] and energies. The main assumption here is that the real GES resembles the GES-like systems in the Auriga simulations.

\section{Conclusions}
\label{sec:conclusions}

Arguably the most significant merger that happened in our Galaxy's history, the GES event, has been studied in great detail utilizing a variety of selection methods and surveys. In this work, we try to understand if we can really pick and choose the way that we select these accreted halo stars and how this affects the inferred progenitor properties. We do this in two parts---first in the observations then in the simulations where we know the \textit{true} GES-like population. The main takeaways from this work are as follows: 


\begin{itemize}
\item{\textbf{In the observations,} given the same APOGEE+Gaia parent sample, we select between 144 (from Jr-Lz) and 1,279 (from eccentricity) GES stars depending on the selection method. These GES samples when projected onto the other selection diagrams show varying levels of agreement and contamination (Figure \ref{fig:ges_chem_4panels}).}
\item{The level of contamination is further supported by the MDFs of the GES samples: the E-Lz, eccentricity, and action diamond methods all show multimodal distributions, with a higher [Fe/H] peak corresponding to in-situ stars. On the other hand the [Mg/Mn]-[Al/Fe] and Jr-Lz methods have more well-behaved, normal distributions, indicating purer selections. From Gaussian mixture modeling, the GES MDF peaks somewhere between $\rm -1.28 < [Fe/H] < -1.18$ dex (Figure \ref{fig:MDFs_4selection}). }
\item{The range in [Fe/H] naturally gives a range in the \mstar~estimate from the MZR that are different by a factor of 2 between the different selections. However, more importantly than the selection method, the adopted MZR is the more sensitive factor that changes the estimated \mstar~by an order of magnitude (Table \ref{tab:MDF}). Based on the [Mg/Mn]-[Al/Fe] MDF in the observations, GES has a lower limit on the \mstar~of $2.37 \times 10^{8}$ $\rm M_{\odot}$ and an upper limit set by the total stellar halo mass (e.g., $\sim$1.4$ \times~10^{9}$ $\rm M_{\odot}$, \citealt{deason19}). The \mstar~based on the average of all five methods in the observations using the $z=0$ relation is $\sim 4 \times 10^{8}$ $\rm M_{\odot}$.}
\item{\textbf{In the simulations,} we found that the Jr-Lz method is best in selecting the purest GES sample at 33\% purity and the eccentricity method is best in selecting the most complete GES sample at 62\% completeness (Figure \ref{fig:ges_panels_sims} and Table \ref{tab:pure_comp}). }
\item{The observed \feh~for GES is generally higher compared to the \feh~from the total GES population due to the presence of a negative metallicity gradient in the progenitor (Figures \ref{fig:mdf_sims_gse_ins} and \ref{fig:mdf_1to1_line}). }
\item{Using the MZR from the destroyed satellites in the Auriga to estimate \mstar~(Figure \ref{fig:mzr_sims}), we found that the eccentricity method tends to overestimate the most by a factor of 2.6, followed by the Jr-Lz method by a factor of 1.5. The action diamond gives the most accurate \mstarest~while the E-Lz method underestimates the true \mstar~of GES by a factor of 0.7. Nonetheless, they all are able to get close to the true \mstar~depending on the Auriga halo (Figure \ref{fig:mratio_violin}). }
\item{\textbf{Bringing together the observations and simulations,} we found that the GES and in-situ stars in the simulations have distinct [Fe/H] trends in the eccentricity vs $\rm R_{GC}$ space i.e., the GES stars tend to have higher [Fe/H] at higher eccentricities while the opposite trend exists for the in-situ material. This is qualitatively reproduced by the [Mg/Mn]-[Al/Fe] and E-Lz methods in the observations (Figure \ref{fig:ecc_v_R}).}
\item{Lastly, we are able to estimate a total mass for GES  ($10^{10.5 - 11.1}~M_{\odot}$) using only the relationship between the metallicity gradient and energy of the GES remnant, and the underlying distribution of GES total masses that drive this relation in the Auriga simulations (Figure \ref{fig:fehgrad_esink_mass}). This is very encouraging as it does not rely on the MZR and stellar mass-halo mass relation. } 
\end{itemize}

Although we have ventured to find the one selection method to rule them all, we have found that each method is uniquely good for estimating some progenitor properties, but not \textit{all} across the board. Perhaps unsurprisingly, the selection method should instead be driven by the science question. In this work, we found that the chemical selection is best in estimating the \mstar~based on the MDF in the observations while the action diamond method is best for an \mstar~estimate in the simulations \footnote{We note here again that the same elements use in the observations are not present in the simulations so we are not able to directly compare.}. The Jr-Lz method gives the purest sample, while the eccentricity method is the most complete. In the future, especially with dedicated surveys aimed at observing fainter targets than APOGEE (e.g., DESI, WEAVE), we can extend our questions into the fuller picture and nature of GES as we will have access to its more distant and presumably least bound stars. 

\section*{Acknowledgements}
AC and AD acknowledge support from the Science and
Technology Facilities Council (STFC) [grant number
ST/T000244/1] and the Leverhulme Trust. AD is supported by a Royal Society
University Research Fellowship. RG acknowledges financial support from the Spanish Ministry of Science and Innovation (MICINN) through the Spanish State Research Agency, under the Severo Ochoa Program 2020-2023 (CEX2019-000920-S), and support from an STFC Ernest Rutherford Fellowship (ST/W003643/1). AF is supported by a UKRI Future Leaders Fellowship (grant no. MR/T042362/1).  

This work used the DiRAC@Durham facility managed by the Institute for Computational Cosmology on behalf of the STFC DiRAC HPC Facility (www.dirac.ac.uk). The equipment was funded by BEIS capital funding via STFC capital grants ST/K00042X/1, ST/P002293/1, ST/R002371/1 and ST/S002502/1, Durham University and STFC operations grant ST/R000832/1. DiRAC is part of the National e-Infrastructure.

\section{Data Availability}
The APOGEE DR17 data used in this article are available at \url{https://www.sdss.org/dr17}. The Gaia data used in this article are available at  \url{https://gea.esac.esa.int/archive/}. Other data used in this article can be made available upon reasonable request to the corresponding authors.



\bibliographystyle{mnras}
\bibliography{aaa_blob} 





\bsp	
\label{lastpage}
\end{document}